\documentclass[preprint,aps,showpacs]{revtex4-1}
\usepackage{amsmath,amssymb,amsfonts,dcolumn,color,graphicx,graphics,latexsym,placeins,epsfig,tikz}
\usetikzlibrary{arrows,shapes}
\usepackage{subfigure,rotating,hyperref}

\newcommand{\be}{\begin{equation}}
\newcommand{\ee}{\end{equation}}
\newcommand{\ba}{\begin{eqnarray}}
\newcommand{\ea}{\end{eqnarray}}

\begin{document}

\title{\Large \bf Evolution of geodesic congruences in a gravitationally collapsing scalar field background}

\author{Rajibul Shaikh}
\email{rajibulshaikh@cts.iitkgp.ernet.in}
\author{Sayan Kar}
\email{sayan@phy.iitkgp.ernet.in}
\author{Anirvan DasGupta}
\email{anir@mech.iitkgp.ernet.in}
\affiliation{${}^{*}$ Centre for Theoretical Studies, Indian Institute of Technology Kharagpur, Kharagpur 721 302, India.}
\affiliation{${}^{\dagger}$ Department of Physics, {\it and} Centre for Theoretical Studies \\ Indian Institute of Technology Kharagpur, Kharagpur 721 302, India.}
\affiliation{${}^{\ddagger}$ Department of Mechanical Engineering, {\it and}
Centre for Theoretical Studies \\ Indian Institute of Technology Kharagpur, Kharagpur
721 302, India.}

\begin{abstract}
\noindent The evolution of timelike geodesic congruences in 
a spherically symmetric, nonstatic, inhomogeneous spacetime 
representing gravitational collapse of a massless scalar field is 
studied.
We delineate how  initial values of the
expansion, rotation and shear of a congruence, 
as well as the spacetime curvature, 
influence the global behavior and focusing properties 
of a family of trajectories. 
Under specific conditions, 
the expansion scalar is shown to
exhibit a finite jump (from negative to positive value) before focusing
eventually occurs. This nonmonotonic behavior of the expansion, 
observed in our numerical work, 
is successfully explained through an analysis of the equation for the
expansion. Finally, we bring out 
the role of the metric parameters (related to
nonstaticity and spatial inhomogeneity), in 
shaping the overall behavior of geodesic congruences.
\end{abstract}


\maketitle

\section{Introduction}
\label{introduction}
\noindent In general relativity (GR), the behavior of a family of test 
particles (described by a nonspacelike geodesic congruence), in a given 
spacetime background, is analyzed by studying the evolution 
of three kinematic variables-- expansion, shear, and rotation (ESR). 
The evolution of the ESR along the congruence, is governed by the Raychaudhuri equations \cite{akr,hawk1,wald}. It is well known that the 
Raychaudhuri equations 
play a crucial role in the context of the Penrose-Hawking 
singularity theorems \cite{penrose,hawk2}. 

\noindent The structure and geometric features of a given spacetime 
(encoded in the metric $g_{ij}$ and its derivatives) must necessarily be 
reflected in the evolution of the kinematic variables that characterize a  
geodesic congruence. Apart from initial conditions on the
kinematic variables, geometric quantities (e.g., the Ricci tensor, Ricci scalar,
and the Weyl tensor) that appear in the Raychaudhuri equations,
also influence the evolution of the congruence. 
Thus, knowing the kinematics of geodesic congruences surely helps 
in probing the spacetime geometry. In addition, we know 
that, observationally, one of the ways to verify the existence and nature of
a given spacetime geometry (and the gravitational field it represents)
is through a study of trajectories.

\noindent The evolution of the kinematic variables of 
timelike geodesic congruences has been 
extensively studied in various spacetime backgrounds in the
recent past \cite{ADG1,SG,ADG2}. 
However, much of this earlier work has been in spacetimes 
that are static. Studies on the evolution of geodesic congruences in spherically symmetric, inhomogeneous, nonstatic spacetimes such as those representing 
gravitational collapse 
have not been looked at yet. One might expect that issues specific to
collapsing scenarios, which include the formation of 
singularities, apparent horizons, conjectures, theorems related to 
singularities etc., may be understood through 
such studies.  

\noindent An important and well-known result that follows from the
Raychaudhuri equation for the expansion is that of geodesic
focusing. A congruence has a focal point if all geodesics in the family 
converge and meet there, at a finite value of the affine parameter. 
Geodesic focusing may be completely benign;
i.e., a focal point may not be a curvature singularity, but the
geodesics in a family intersect at such a focal point, thereby 
defining the notion of a congruence singularity. On the other hand,
curvature singularities must always be focal points
of a geodesic congruence. 

\noindent Even though much has been said and proved 
about geodesic focusing, questions do remain. In particular, the
questions we wish to analyze are largely related to detailed
studies from which, some surprises may emerge. Some such questions are as follows:

\noindent $\bullet$ How and in what ways do the 
{\em initial values} of ESR influence the
occurrence of focusing/defocusing? 

\noindent $\bullet$ What are the {\em precise} roles of spacetime
curvature and other geometric quantities in the evolution of congruences? 

\noindent $\bullet$ Besides monotonic focusing and defocusing, are there any 
other characteristic features in the evolution
of the expansion, shear, or rotation that may be correlated with
properties of a given spacetime geometry?

\noindent By adopting methods developed in some of our earlier papers  
\cite{ADG1,SG,ADG2}, in this article, we try to answer some of these questions in the
context of a class of spherically symmetric, nonstatic, inhomogeneous 
spacetimes representing gravitational collapse.

\noindent Of course, in order to proceed with our studies, 
we need a spacetime line element representing 
gravitational collapse. Among many available models, we 
choose an exact solution representing 
a scalar field collapse scenario in (3+1) dimensions \cite{viqar1}. 
Our choice is essentially governed by an available exact, spherically symmetric, nonstatic,
inhomogeneous
solution that is reasonably simple in its line element structure. 
However, note that in this solution the scalar field is
all pervading (exists for all $r$ and $t$) and the collapse scenario 
is somewhat different from the
usual pressureless dust ball collapse (Oppenheimer-Snyder) 
or even the spherisymmetric  collapse of a perfect fluid with pressure.

\noindent The above-mentioned spacetime, along with its geodesic structure 
is discussed 
in Sec. \ref{spacetime}. In Sec. \ref{Raychaudhuri_eqn}, we briefly review the 
derivation of the Raychaudhuri equations for geodesic congruences. 
In Sec. \ref{ESR_evolution_timelike}, we solve the 
ESR evolution equations 
(along with the geodesic equations) numerically and 
bring out certain aspects of the evolution kinematics of the congruence.
An interesting behavior of the expansion under specific conditions
is explained in a general context in Sec. \ref{analysis}. The role of the metric parameters related to
nonstaticity and spatial inhomogeneity in controlling and characterizing
the evolution of congruences is studied in Sec. \ref{metric_parameter}.
Finally, Sec. \ref{summary_conclusion} summarizes our results and suggests
future avenues of work.

\section{\bf Exact solution for scalar field collapse}
\label{spacetime}
\noindent We begin by writing the line element representing 
the gravitational collapse of a massless scalar field minimally coupled to 
(3+1)-dimensional gravity \cite{viqar1}
\begin{equation}
ds^2=(at+b)\left[-f^2(r)dt^2+f^{-2}(r)dr^2\right]+R^2(r,t)(d\psi^2+\sin^2\psi d\phi^2)
\label{eq:metric}
\end{equation}
where
\begin{equation*}
f^2(r)=\left(1-\frac{2c}{r}\right)^{\alpha}
\end{equation*}
\begin{equation*}
R^2(r,t)=(at+b)r^2 \left(1-\frac{2c}{r}\right)^{1-\alpha}
\end{equation*}
and the scalar field profile is given as,
\begin{equation*}
\Phi(r,t)=\pm \frac{1}{4\sqrt{\pi}}\ln\left[d\left(1-\frac{2c}{r}\right)^{\frac{\alpha}{\sqrt{3}}}(at+b)^{\sqrt{3}}\right]
\end{equation*}
where $a$, $b$, $c$, $d$ are constants and $\alpha=\pm\frac{\sqrt{3}}{2}$. $R(r,t)$ is the area radius. 

\begin{figure}[ht]
\centering
\subfigure[$\alpha=-\frac{\sqrt{3}}{2}$]{\includegraphics[scale=0.93]{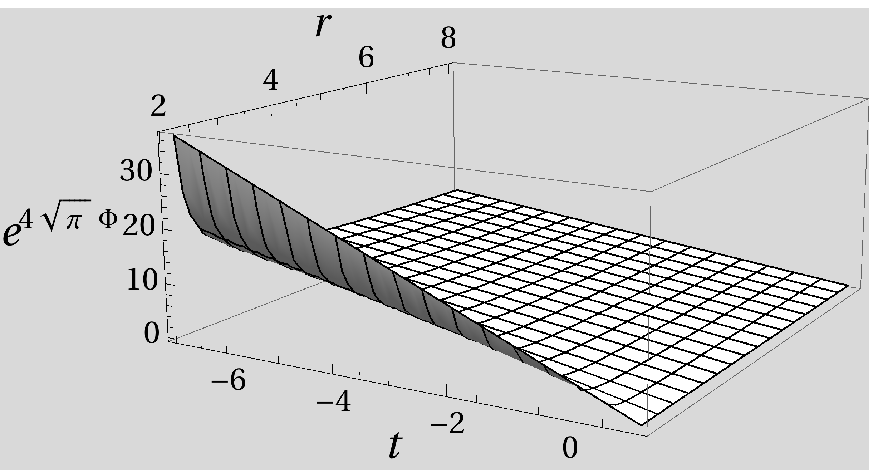}\label{fig:profile_1}}
\subfigure[$\alpha=\frac{\sqrt{3}}{2}$]{\includegraphics[scale=0.75]{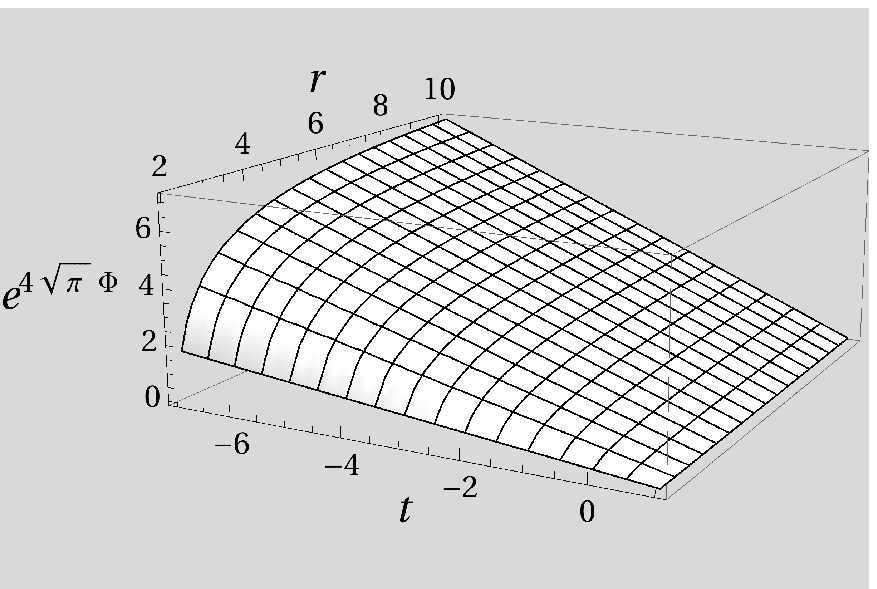}\label{fig:profile_2}}
\caption{Plots for the scalar field profile for $a=-1.0,b=1.0,c=1.0,d=1.0$.}
\label{fig:profile}
\end{figure}

\noindent The above line element and the scalar field constitute solutions of the Einstein field equations given as
\begin{equation}
R_{\alpha \beta} =  8\pi \partial_\alpha \Phi \partial_\beta \Phi
\end{equation}
The plots for the scalar field profile are shown in Fig. \ref{fig:profile}. 
The curvature scalar for the metric is
\begin{equation*}
{\cal R}=\frac{12ca^2(r-c)-3a^2r^2}{2r^2(at+b)^3}\left(1-\frac{2c}{r}\right)^{-2-\alpha}+\frac{2c^2(1-\alpha^2)}{(at+b)r^4}\left(1-\frac{2c}{r}\right)^{-2+\alpha}
\end{equation*}
It should be noted that curvature singularities are present at $r=2c$ and at 
$t=-b/a$ for both values of $\alpha$. Depending on the constants $a$, $b$ and $c$, the metric (\ref{eq:metric}) represents different types of spacetimes. 
Details about these solutions are available in \cite{viqar1}. The constant $c$ is related to the central inhomogeneity of the matter distribution. In the 
limit $c\to 0$, the $r$ dependence of the metric is removed and the spacetime 
becomes homogeneous \cite{clifton,faraoni}. $a$ is the nonstaticity
parameter. $a<0$ and $a>0$ represent 
black-hole-like and white-hole-like solutions, respectively. We consider 
the case $a<0$. The apparent horizon is described by $g^{\alpha\beta}\partial_\alpha R\partial_\beta R=0$. The time of formation of the
apparent horizon at a Lagrangian coordinate $r$ is given by
\begin{equation}
t_{ah}(r)=-\frac{b}{a} \pm \frac{r^2}{2}(1-\frac{2c}{r})^{1-\alpha}[r-c(1+\alpha)]^{-1}
\end{equation}
where plus (minus) sign is for $a>0$ ($a<0$). For $a<0$, the two-sphere labeled by $r$ gets trapped ($g^{\alpha\beta}\partial_\alpha R\partial_\beta R<0$) for $t>t_{ah}(r)$.

\noindent The geodesics in the equatorial section $(\psi=\pi/2)$ of the 
collapsing spacetime are governed by
\begin{equation}
\dot{\phi}=\frac{L}{R^2(z,t)}
\label{eq:geodesic_phi}
\end{equation}
\begin{equation}
\dot{z}=f(z)\sqrt{f^2(z)\dot{t}^2+\frac{sR^2(z,t)-L^2}{R^2(z,t)(at+b)}}
\label{eq:geodesic_z}
\end{equation}
\begin{equation}
\ddot{t}+\frac{a}{(at+b)}\dot{t}^2+2\frac{f'(z)}{f(z)}\dot{t}\dot{z}+\frac{as}{2f^2(z)(at+b)^2}=0
\label{eq:geodesic_t}
\end{equation}
where $L$ is an integration constant representing angular momentum. 
In obtaining Eqs. (\ref{eq:geodesic_phi})-(\ref{eq:geodesic_t}), we have used the transformation $r-2c=z$. 
Also, we have used the fact that the velocity vector $u^\alpha=(\dot{t},\dot{z},\dot{\phi})$ satisfies the constraint $u^\alpha u_\alpha=s$, where $s=-1$ for timelike geodesics and $s=0$ for null geodesics.

\section{The Raychaudhuri equations}
\label{Raychaudhuri_eqn}
\noindent Consider a congruence of timelike geodesics in a given spacetime 
background. The geodesic congruence may undergo isotropic expansion, shear, and rotation (ESR). The kinematics of these quantities are investigated in the spacelike 
hypersurface orthogonal to the central geodesic. Therefore, one can define 
a transverse spatial metric $h_{\alpha\beta}$ induced on the spacelike 
hypersurface as
\begin{equation}
h_{\alpha\beta}=g_{\alpha\beta}+u_\alpha u_\beta; \hspace{1.0cm} (\alpha,\beta=0,1,2,3)
\label{eq:transverse_metric}
\end{equation}
where the timelike vector field $u^\alpha$ associated with the congruence is 
tangent to the geodesic at each point and satisfies the timelike constraint 
$u^\alpha u_\alpha=-1$. The transverse metric satisfies 
$u^\alpha h_{\alpha\beta}=0$, implying that $h_{\alpha\beta}$ is orthogonal 
to $u^\alpha$. From the vector field $u^\alpha$, one can define the velocity gradient tensor 
$B_{\alpha\beta}=\nabla_{\beta} u_\alpha$. In four spacetime dimensions, 
the tensor $B_{\alpha\beta}$ can be decomposed into its 
trace, symmetric traceless, and antisymmetric parts as
\begin{equation}
B_{\alpha\beta}=\frac{1}{3}h_{\alpha\beta}\theta+\sigma_{\alpha\beta}+\omega_{\alpha\beta}
\label{eq:bdef}
\end{equation}
where $\theta=B^\alpha_{\;\alpha}$ is the expansion scalar (trace part), 
$\sigma_{\alpha\beta}=\frac{1}{2}(B_{\alpha\beta}+B_{\beta\alpha})-
\frac{1}{3}h_{\alpha\beta}\theta$ the shear tensor (symmetric traceless part) 
and $\omega_{\alpha\beta}=\frac{1}{2}(B_{\alpha\beta}-B_{\beta\alpha})$ the 
rotation tensor (antisymmetric part). By virtue of this construction, the shear and the rotation tensors 
satisfy $h^{\alpha\beta}\sigma_{\alpha\beta}=0$ 
and $h^{\alpha\beta}\omega_{\alpha\beta}=0$. We also have 
$g^{\alpha\beta}\sigma_{\alpha\beta}=0$ and 
$g^{\alpha\beta}\omega_{\alpha\beta}=0$. 
Since $u^\alpha\sigma_{\alpha\beta}=0$ and $u^\alpha\omega_{\alpha\beta}=0$, 
both $\sigma_{\alpha\beta}$ and $\omega_{\alpha\beta}$  are 
purely spatial (i.e., $\sigma^{\alpha\beta}\sigma_{\alpha\beta}>0$ 
and $\omega^{\alpha\beta}\omega_{\alpha\beta}>0$) and lie in the orthogonal 
hypersurface.

\noindent The evolution equation for the spatial tensor $B_{\alpha\beta}$ 
can be written as,
\begin{equation}
u^\gamma \nabla_\gamma B_{\alpha\beta}=-B_{\alpha\gamma}B^\gamma_{\;\beta}+R_{\gamma\beta\alpha\delta}u^\gamma u^\delta
\label{eq:evolution_eqn}
\end{equation}
where $R_{\gamma\beta\alpha\delta}$ is the Riemann tensor. The trace, symmetric traceless, and antisymmetric parts of the equation yield \cite{wald}
\begin{equation}
\frac{d\theta}{d\lambda}+\frac{1}{3}\theta^2+\sigma^2-\omega^2+R_{\alpha\beta}u^\alpha u^\beta=0
\label{eq:expansion_eqn}
\end{equation}
\begin{equation}
u^\gamma \nabla_\gamma \sigma_{\alpha\beta}+\frac{2}{3}\theta\sigma_{\alpha\beta}+\sigma_{\alpha\gamma}\sigma^{\gamma}_{\;\beta}+\omega_{\alpha\gamma}\omega^{\gamma}_{\;\beta}-\frac{1}{3}h_{\alpha\beta}(\sigma^2-\omega^2)-C_{\gamma\beta\alpha\delta} u^\gamma u^\delta-\frac{1}{2}\tilde{R}_{\alpha\beta}=0
\label{eq:shear_eqn}
\end{equation}
\begin{equation}
u^\gamma \nabla_\gamma \omega_{\alpha\beta}+\frac{2}{3}\theta\omega_{\alpha\beta}+\sigma^\gamma_{\;\beta}\omega_{\alpha\gamma}-\sigma^\gamma_{\;\alpha}\omega_{\beta\gamma}=0
\label{eq:rotation_eqn}
\end{equation}
where $\sigma^2=\sigma^{\alpha\beta}\sigma_{\alpha\beta}$, 
$\omega^2=\omega^{\alpha\beta}\omega_{\alpha\beta}$, $\lambda$ is the affine 
parameter, $C_{\gamma\beta\alpha\delta}$ is the Weyl tensor and 
$\tilde{R}_{\alpha\beta}=(h_{\alpha\gamma}h_{\beta\delta}-\frac{1}{3}h_{\alpha\beta}h_{\gamma\delta})R^{\gamma\delta}$ is the transverse trace-free part of 
$R_{\alpha\beta}$. The equation for $\theta$ is a Riccati-type equation, 
and is of considerable interest in the context of the singularity theorems. 
Redefining $\theta=3\frac{\dot{F}}{F}$, one can obtain the following Hill-type 
equation:
\begin{equation}
\frac{d^2F}{d\lambda^2}+\frac{1}{3}(R_{\alpha\beta}u^\alpha u^\beta+\sigma^2-\omega^2)F=0.
\end{equation}
The analysis of focusing ($\theta\rightarrow-\infty$) or defocusing ($\theta\rightarrow\infty$) can be done by investigating the quantity 
$I=R_{\alpha\beta}u^\alpha u^\beta+\sigma^2-\omega^2$. It is clear from (\ref{eq:expansion_eqn}) 
that the sufficient condition for geodesic focusing is $I> 0$. 
Timelike convergence condition requires $R_{\alpha\beta}u^\alpha u^\beta\geq 0$. Thus, from the signs of the various terms in the equation for
the expansion, we can conclude that rotation defies focusing, whereas 
shear assists it.

\noindent In the context of this paper and from the Einstein-scalar equations,
we have
$R_{\alpha\beta} = 8\pi\partial_\alpha \phi \partial_\beta \phi$. 
Therefore,
\begin{equation}
I = \left[\sigma^2 -\omega^2 + 8\pi\left ( \frac{d\phi}{d\lambda}\right )^2\right]
\end{equation}
The third term on the right-hand side above is always positive. Hence, the
positivity of $I$ is crucially dependent on the sign of
$\sigma^2-\omega^2$. As we see, $\sigma^2-\omega^2$ may be positive
or negative over a certain domain of $\lambda$.  Thus,
it is possible that there exists a domain of $\lambda$ where
$I<0$. 
We see in detail later how the fact that $I<0$ 
over a restricted domain of $\lambda$ gives rise to 
a distinct
new feature in the evolution of the expansion scalar $\theta$.  
In particular, we observe that the expansion scalar can 
exhibit a glitch or a jump in its evolution before
eventually focusing. 


\section{Evolution of a timelike geodesic congruence}
\label{ESR_evolution_timelike}

\subsection{Method of solution}

\noindent Equations (\ref{eq:expansion_eqn})-(\ref{eq:rotation_eqn}) are nonlinear coupled, ordinary differential equations. In the absence of analytical solutions, one has to solve this set of equations (along with the geodesic equations) numerically. However, instead of solving Eqs. (\ref{eq:expansion_eqn})-(\ref{eq:rotation_eqn}), it is more convenient to solve Eq. (\ref{eq:evolution_eqn}) and subsequently extract the expansion scalar $\theta$, shear tensor $\sigma_{\alpha\beta}$, and rotation tensor $\omega_{\alpha\beta}$ by taking the trace, symmetric traceless, and antisymmetric parts of $B_{\alpha\beta}$, respectively. The initial condition on $B_{\alpha\beta}$ can be easily constructed from the initial conditions on the ESR variables using (\ref{eq:bdef}). It may be pointed out that initial conditions on the velocity field $u^\alpha$, $\sigma_{\alpha\beta}$ and $\omega_{\alpha\beta}$ must satisfy the orthogonality conditions.

\noindent We first study the kinematic evolution of geodesic congruences for the black-hole-like solution ($a=-1$, $b=1$, and $c=1$). There is a timelike singularity at $r=2$, i.e., at $z=0$, and a spacelike singularity at $t=1$. For the latter case, the range of the time coordinate is $-\infty<t\leq 1$. At time $t=1$, the whole spacetime collapses to the origin $R=0$. Therefore, for geodesics 
beginning at time $t<1$, the spacelike singularity at $t=1$ is a future 
directed singularity.
We consider the geodesic congruence in the equatorial section ($\psi=\pi/2$). Throughout the numerical evaluation, we have kept fixed the initial conditions on $\lbrace x^\alpha(\lambda),u^\alpha(\lambda)\rbrace$. 

\subsection{Evolution of kinematic variables for ${\bf \alpha=-\frac{\sqrt{3}}{2}}$}

\begin{figure}[ht]
\centering
\subfigure[$\sigma_0^2=0$]{\includegraphics[scale=0.84]{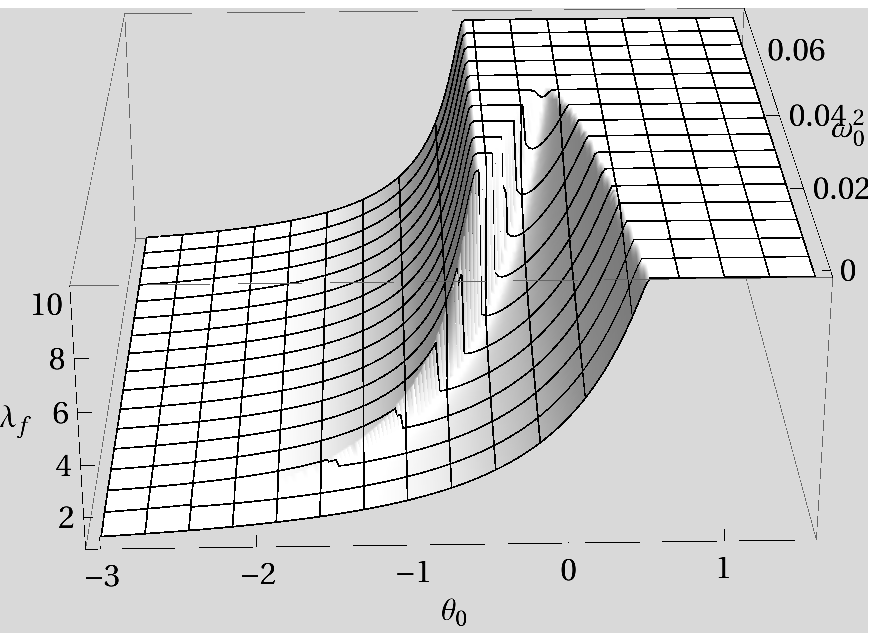}\label{fig:negative_alpha_3D_1}}
\subfigure[$\omega_0^2=0$]{\includegraphics[scale=0.765]{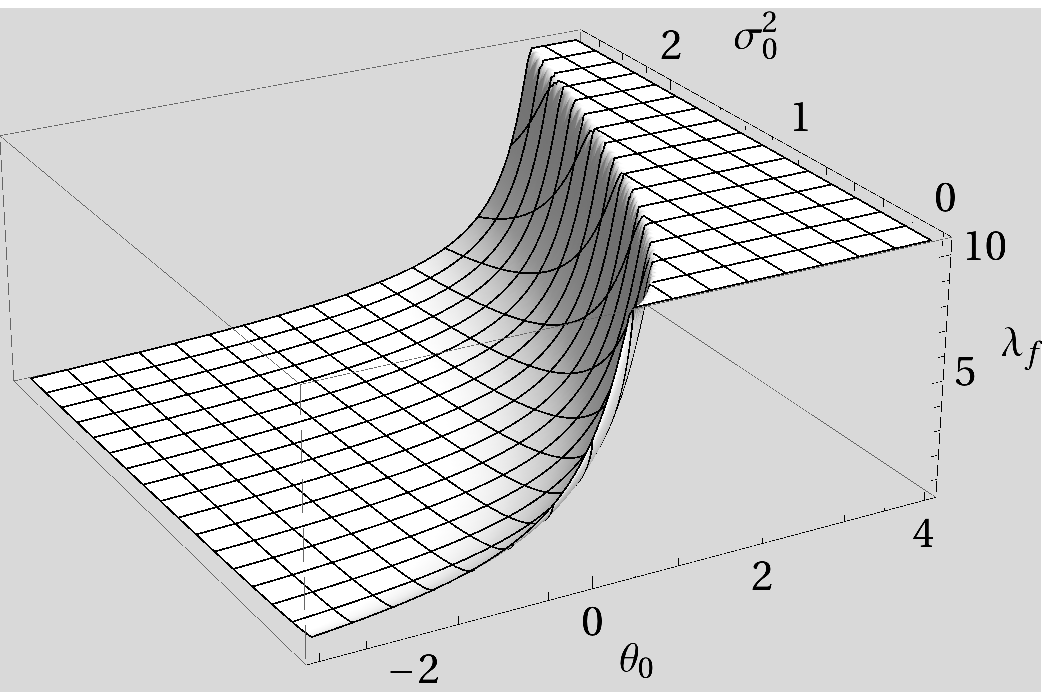}\label{fig:negative_alpha_3D_2}}
\subfigure[$\theta_0=0$]{\includegraphics[scale=0.895]{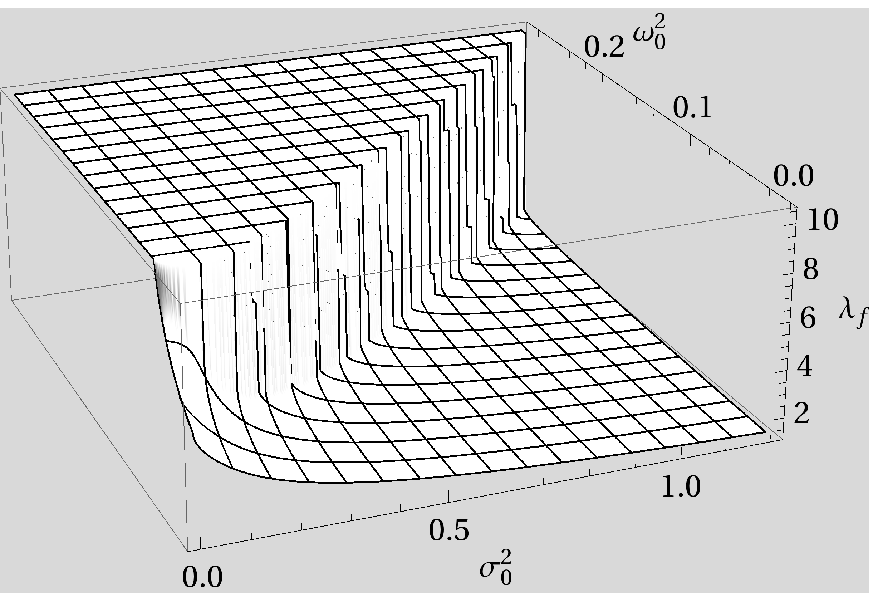}\label{fig:negative_alpha_3D_3}}
\caption{Plots showing the dependence of the focusing affine parameter $\lambda_f$ on the initial ESR for $\alpha=-\frac{\sqrt{3}}{2}$. Here the initial conditions are $t(0)=-7.0$, $z(0)=1.0$, $\phi(0)=0$, $\dot{t}(0)=0.25$, and $L=1.0$.}
\label{fig:negative_alpha_3D}
\end{figure}
\begin{figure}[ht]
\centering
\subfigure[]{\includegraphics[scale=0.75]{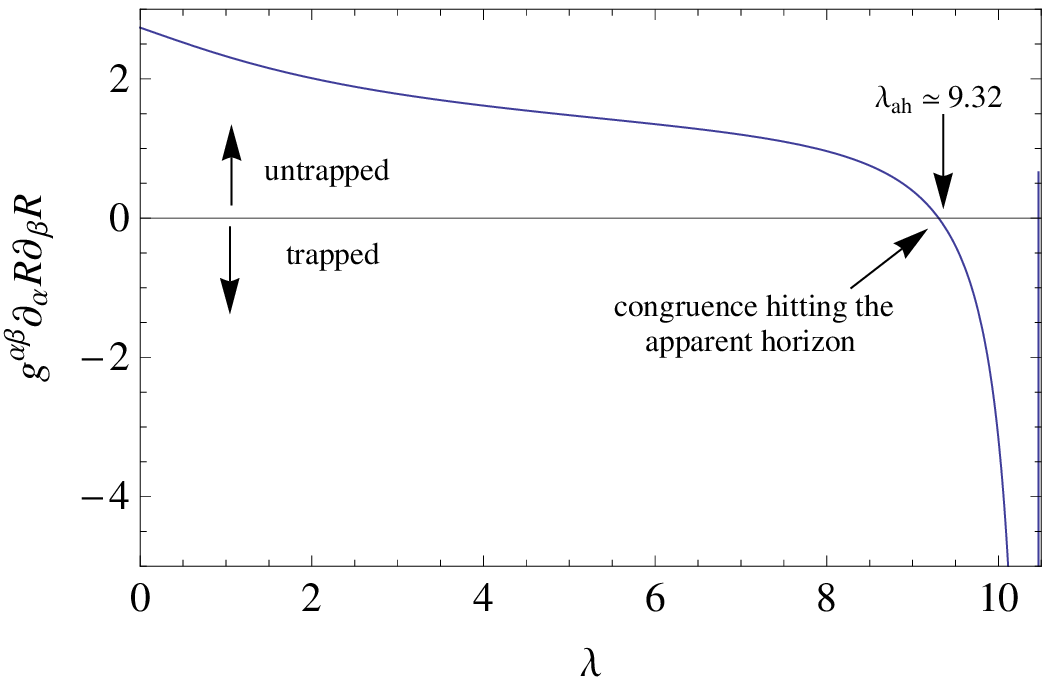}\label{fig:negative_alpha_AH}}
\subfigure[]{\includegraphics[scale=0.74]{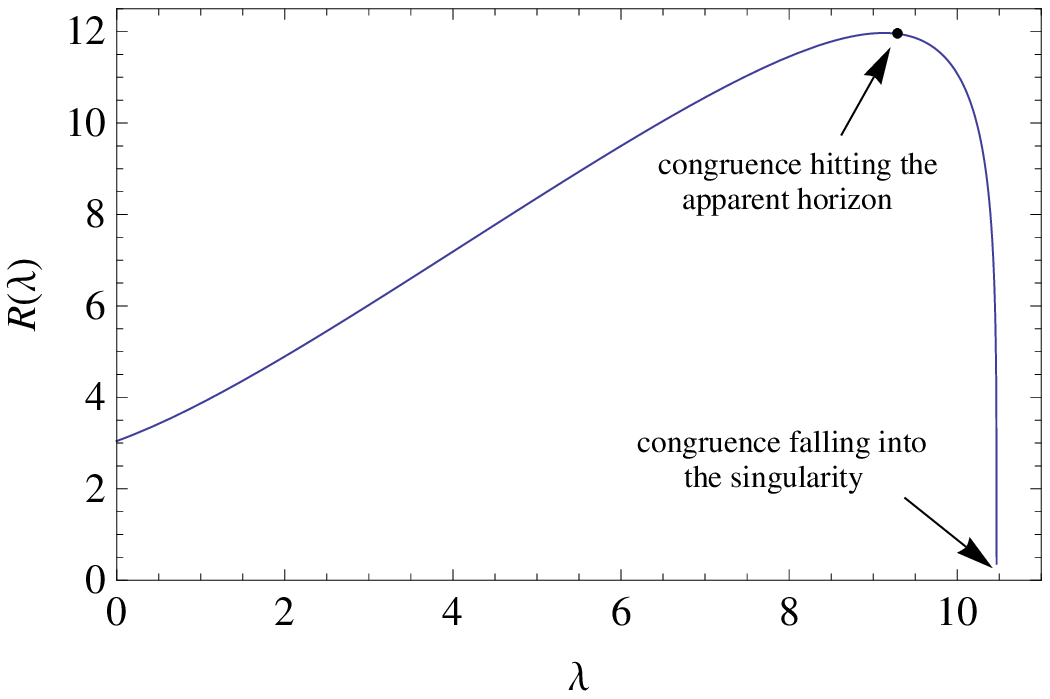}\label{fig:negative_alpha_R}}
\caption{Plots of variation of (a) $g^{\alpha\beta} \partial_\alpha R \partial_\beta R$ and 
(b) area radius $R$ along the geodesic congruence for $\alpha=-\frac{\sqrt{3}}{2}$.}
\label{fig:negative_alpha_AH_R}
\end{figure}

\noindent From the numerical evaluations, with different initial conditions, it is found that 
the congruence always exhibits focusing. We denote $\lambda_f$ as the value of 
the
affine parameter at which focusing takes place. The dependence of $\lambda_f$ on the initial
conditions of the ESR variables is presented in Fig. \ref{fig:negative_alpha_3D}. 
The value $\lambda_f=10.46$ corresponds to the time $t=1$ when the geodesic 
congruence encounters a future spacelike singularity. 
Figure \ref{fig:negative_alpha_AH_R} shows the variations of 
$g^{\alpha\beta}\partial_\alpha R \partial_\beta R$ and the area radius $R$ 
along the geodesic congruence with identically vanishing shear and rotation. It is clear that, as $\lambda \to 10.46$, the area radius $R$ tends 
to zero. Therefore, as $\lambda \to 10.46$, 
the congruence falls into the spacelike singularity. As observed in Fig.~\ref{fig:negative_alpha_AH_R}, the congruence 
encounters the apparent horizon at $\lambda_{ah}=9.32$, after which it
gets trapped ($g^{\alpha\beta}\partial_\alpha R \partial_\beta R<0$)
 in the trapping region formed in the spacetime. The area radius $R$ initially 
increases untill the congruence hits the apparent horizon and is subsequently 
subsumed in the trapped region before it falls into the singularity. 
Therefore, the congruence gets trapped before it falls into the singularity. 
This is because of the fact that, in the given collapsing spacetime, a two-sphere labelled by $r$ becomes trapped before it becomes singular, i.e, 
before it collapses to the origin $R=0$. The plateau-top region in Fig. \ref{fig:negative_alpha_3D} corresponds to the 
time (in terms of the affine parameter $\lambda$) of formation of the spacetime singularity. 
We now discuss the effect of initial ESR, as well as the 
curvature, on the focusing time.
\begin{figure}[ht]
\centering
\includegraphics[scale=0.99]{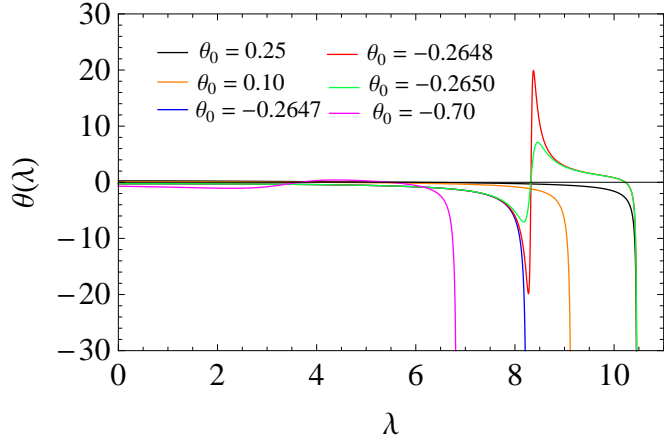}
\caption{Plot of variation of $\theta(\lambda)$ for different initial expansion with $\sigma^2_0=0.0$ and $\omega^2_0=0.04$.}
\label{fig:negative_alpha_expansion_1}
\end{figure}
\begin{figure}[ht]
\centering
\subfigure[$\theta_0=0.25$]{\includegraphics[scale=0.79]{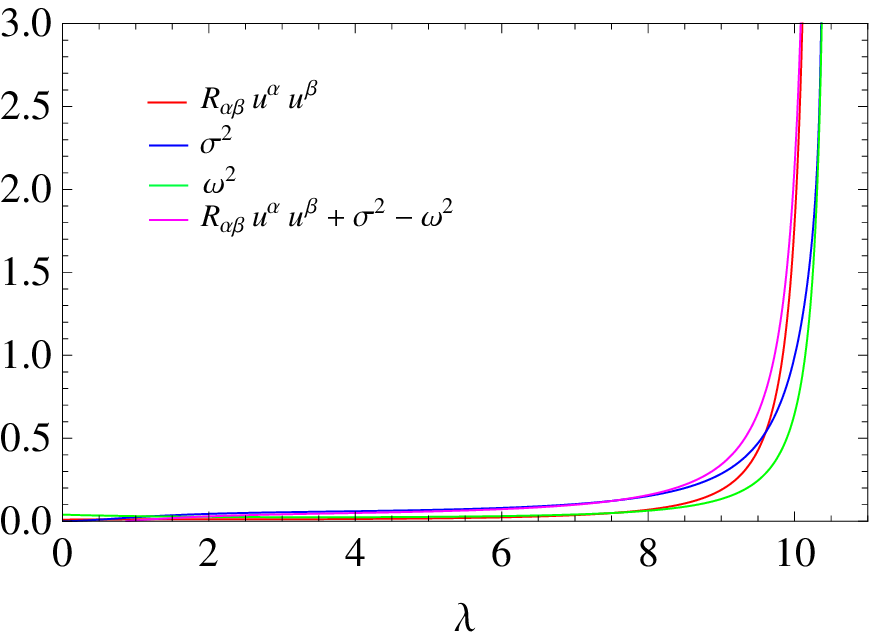}\label{fig:negative_alpha_expansion_2a}}\hspace{0.2cm}
\subfigure[$\theta_0=0.10$]{\includegraphics[scale=0.8]{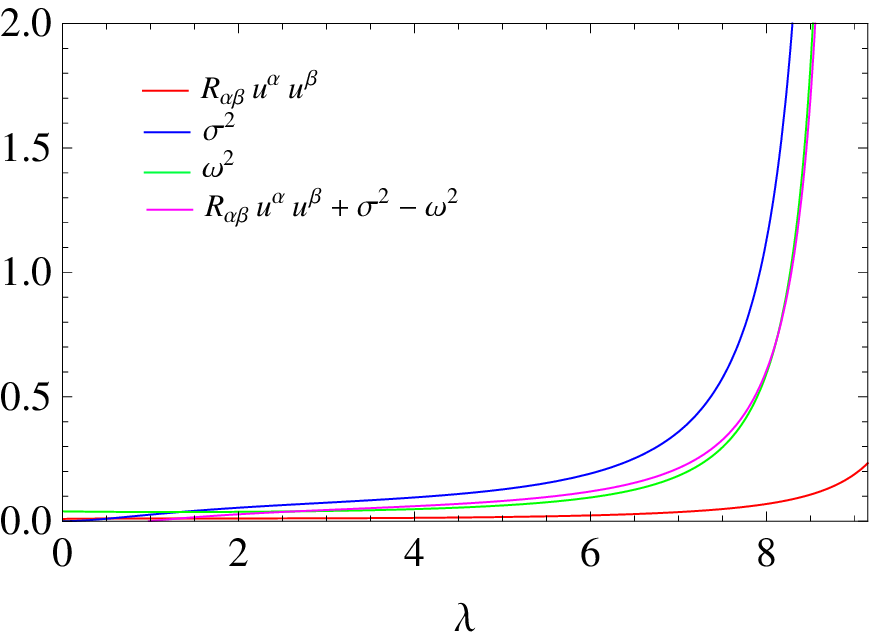}\label{fig:negative_alpha_expansion_2b}}
\subfigure[$\theta_0=-0.2647$]{\includegraphics[scale=0.79]{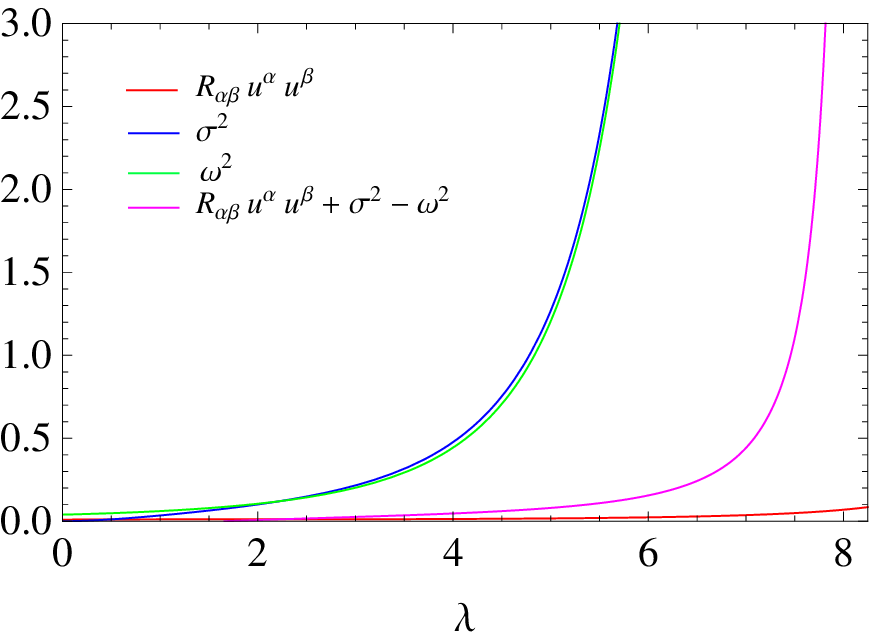}\label{fig:negative_alpha_expansion_2c}}\hspace{0.2cm}
\subfigure[$\theta_0=-0.2648$]{\includegraphics[scale=0.85]{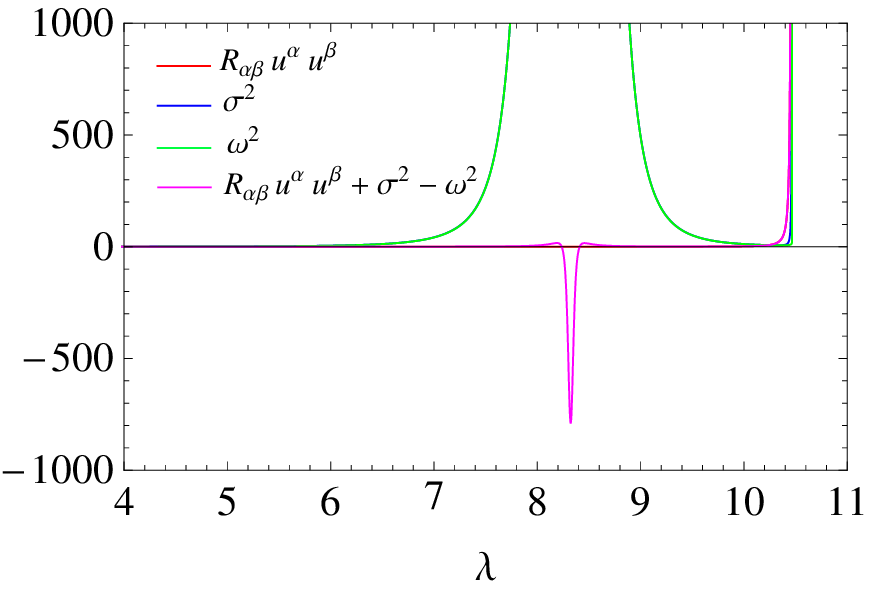}\label{fig:negative_alpha_expansion_2d}}
\subfigure[$\theta_0=-0.2650$]{\includegraphics[scale=0.84]{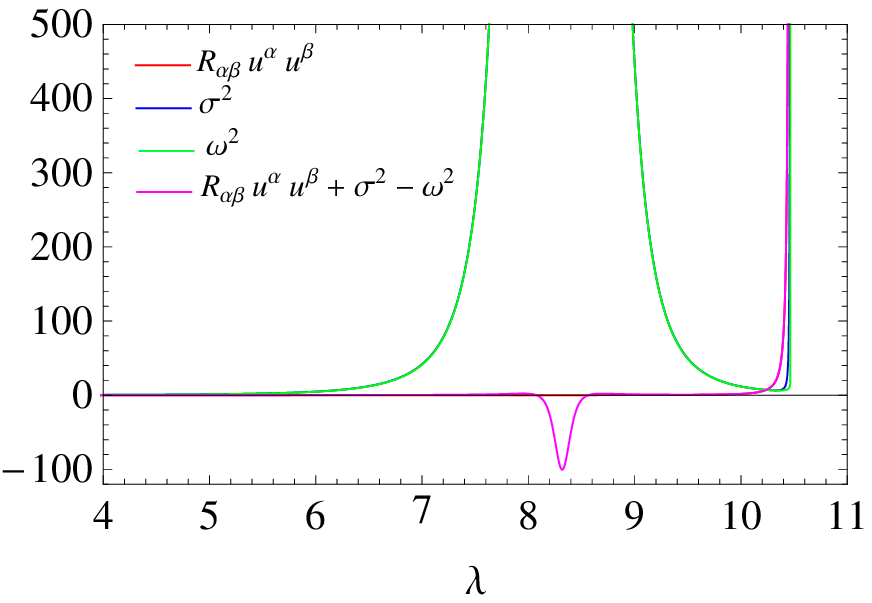}\label{fig:negative_alpha_expansion_2e}}\hspace{0.2cm}
\subfigure[$\theta_0=-0.70$]{\includegraphics[scale=0.8]{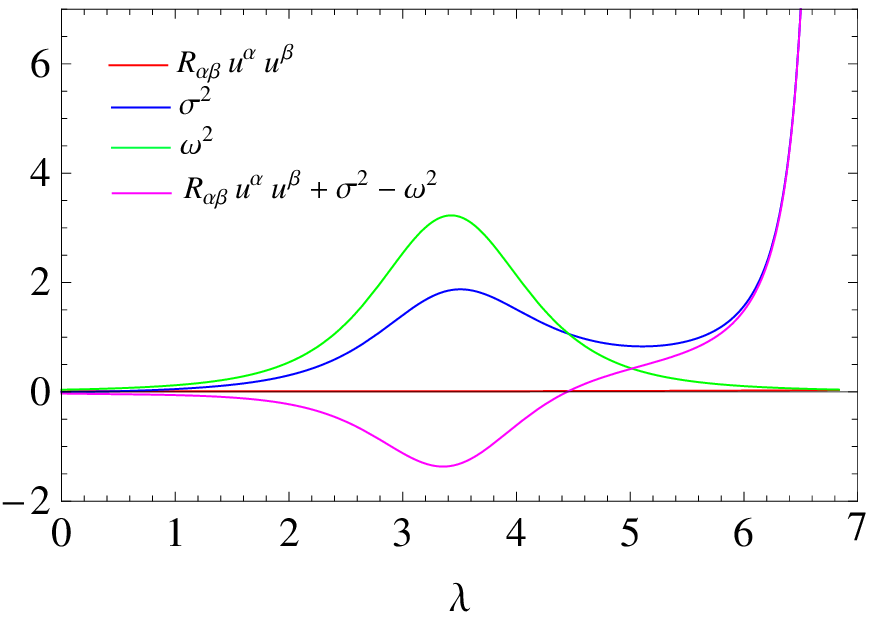}\label{fig:negative_alpha_expansion_2f}}
\caption{Plots of $R_{\alpha\beta}u^\alpha u^\beta$, $\sigma^2$, $\omega^2$ and $I$ ($=R_{\alpha\beta}u^\alpha u^\beta+\sigma^2-\omega^2$) for the initial values of the ESR variables corresponding to Fig. \ref{fig:negative_alpha_expansion_1}}
\label{fig:negative_alpha_expansion_2}
\end{figure}

\subsubsection{Effect of initial expansion and shear on focusing}

\noindent Generally, the effect of increasing (decreasing) the initial 
expansion ($\theta_0$) is to delay (prepone) the focusing, but for $\omega_0$ in a certain range, Fig. \ref{fig:negative_alpha_3D_1} shows a 
peculiar nonmonotonic dependence of $\lambda_f$ on $\theta_0$. For $\omega_0^2$ lying 
between two values $\omega_{c1}^2$ and $\omega_{c2}^2$ ($\omega_{c1}^2<\omega_{c2}^2$), with decreasing $\theta_0$, $\lambda_f$ decreases up to a certain 
initial expansion $\theta_c$; below $\theta_c$, $\lambda_f$ increases suddenly 
and then starts decreasing again. Notice that the value of $\theta_c$ depends 
on $\omega_0^2$.

\noindent To understand the above-mentioned peculiar behavior for 
$\omega_{c1}^2<\omega_0^2<\omega_{c2}^2$, we consider the initial conditions 
corresponding to the section $\omega_0^2=0.04$ in Fig. \ref{fig:negative_alpha_3D_1} and plot 
the evolution of the expansion $\theta(\lambda)$ in Fig. \ref{fig:negative_alpha_expansion_1}. 
It is observed that untill $\theta_0=-0.2647$, $\lambda_f$ decreases with decreasing $\theta_0$. However, between 
$\theta_0=-0.2647$ and $\theta_0=-0.2648$, a transition (sudden change) 
in focusing time is noted; focusing is delayed. Therefore, for 
$\omega^2_0=0.04$, $\theta_{c}$ lies between $-0.2648$ and $-0.2647$. 
In Fig. \ref{fig:negative_alpha_expansion_2}, we investigate
this peculiar behavior in the focusing time by plotting
the corresponding evolution of the scalars $R_{\alpha\beta}u^\alpha u^\beta$, $\sigma^2$, $\omega^2$, and 
$I=R_{\alpha\beta}u^\alpha u^\beta+\sigma^2-\omega^2$.
It is observed that, for $\theta_0=0.25$, $I$ diverges mainly because of the curvature term 
$R_{\alpha\beta}u^\alpha u^\beta$, and the divergence takes place as 
$\lambda\rightarrow 10.46$, implying that the focusing takes place due to 
the singularity formation [Fig. \ref{fig:negative_alpha_expansion_2a}]. 
For $\theta_0=0.1$ or $-0.2647$, $I$ diverges because of the term 
$\sigma^2$, and divergence takes place much before the singularity 
formation, implying that the focusing takes place due to divergence of 
shear [Figs. \ref{fig:negative_alpha_expansion_2b} and \ref{fig:negative_alpha_expansion_2c}]. 
For $\theta_0=-0.2648$, initially $\sigma^2$ and $\omega^2$ 
almost cancel each other in the expression of $I$ [Fig. \ref{fig:negative_alpha_expansion_2d}].
However, midway during the evolution, over a short period, $\omega^2$ dominates over $\sigma^2$ making $I<0$ 
(congruence starts defocusing). This induces a sharp transition/jump (from negative to positive) in the evolution 
of $\theta$, as shown in Fig.~\ref{fig:negative_alpha_expansion_1}. 
As the evolution proceeds further, $I$ again becomes positive and diverges because of the curvature term 
$R_{\alpha\beta}u^\alpha u^\beta$. Therefore, the focusing is delayed 
(see Fig. \ref{fig:negative_alpha_expansion_1}) because of the dominance of 
rotation over shear, midway during the evolution. However, complete 
defocusing does not take place because $R_{\alpha\beta}u^\alpha u^\beta$ 
diverges as the evolution proceeds toward the singularity formation time 
$t=1.0$. The amplitude of the jump in the evolution of the expansion scalar gets smaller as one makes 
the initial expansion $\theta_0$ more negative (Fig.~\ref{fig:negative_alpha_expansion_1}). 
It may again be noted that, for the case $\theta_0=-0.70$, focusing takes place entirely due to the 
divergence of shear [Fig. \ref{fig:negative_alpha_expansion_2f}], and
the curvature singularity has no role in the focusing.

\noindent The $\theta_0=constant$ sections of Fig. \ref{fig:negative_alpha_3D_2} indicate that, 
with increasing $\sigma_0^2$, focusing time decreases monotonically; i.e., initial shear always helps in focusing.
\subsubsection{Effect of initial rotation on focusing}
\begin{figure}[ht]
\centering
\includegraphics[scale=0.90]{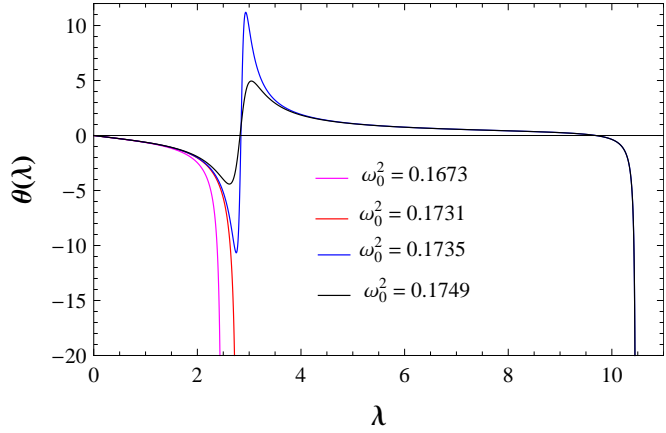}
\caption{Figures showing the role of initial rotation $\omega^2_0$ in focusing of the congruence, for $\theta_0=0.0,\sigma^2_0=0.75$.}
\label{fig:negative_alpha_rotation_1}
\end{figure}

\begin{figure}[ht]
\centering
\subfigure[$\omega^2_0=0.1673$]{\includegraphics[scale=0.95]{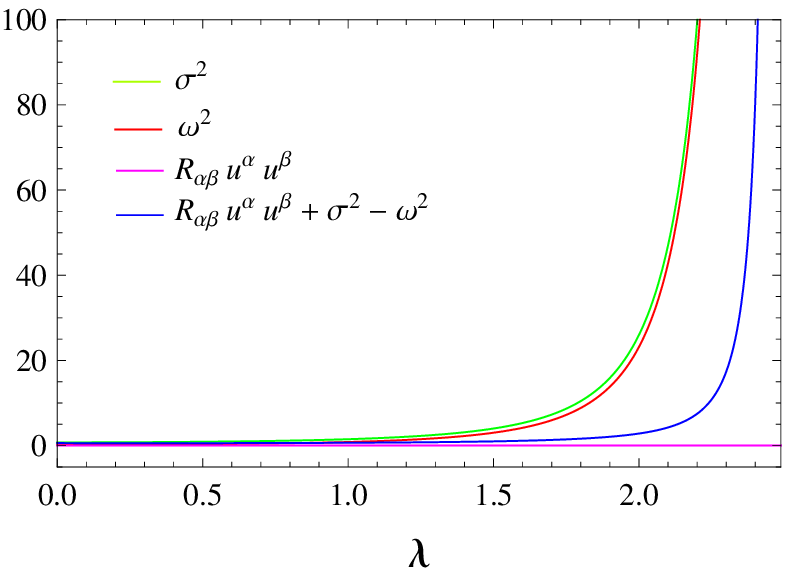}\label{fig:negative_alpha_rotation_2a}}
\subfigure[$\omega^2_0=0.1731$]{\includegraphics[scale=0.95]{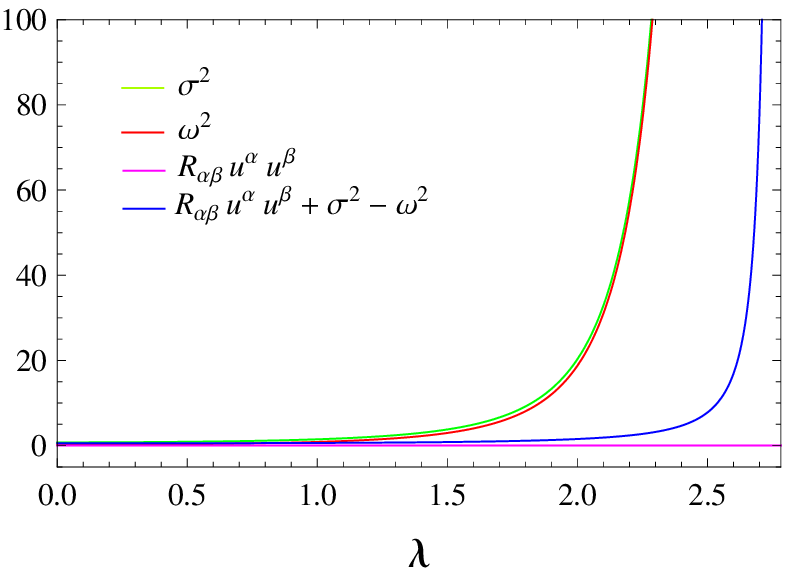}\label{fig:negative_alpha_rotation_2b}}
\subfigure[$\omega^2_0=0.1735$]{\includegraphics[scale=0.99]{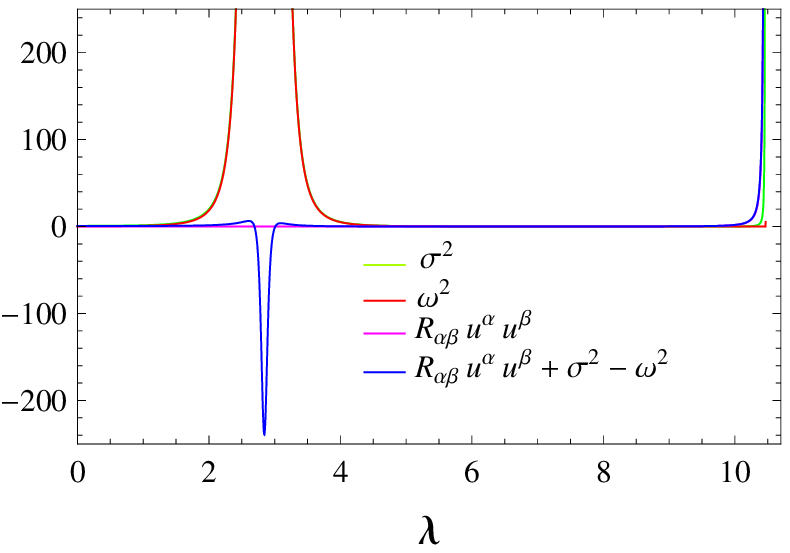}\label{fig:negative_alpha_rotation_2c}}
\subfigure[$\omega^2_0=0.1749$]{\includegraphics[scale=0.95]{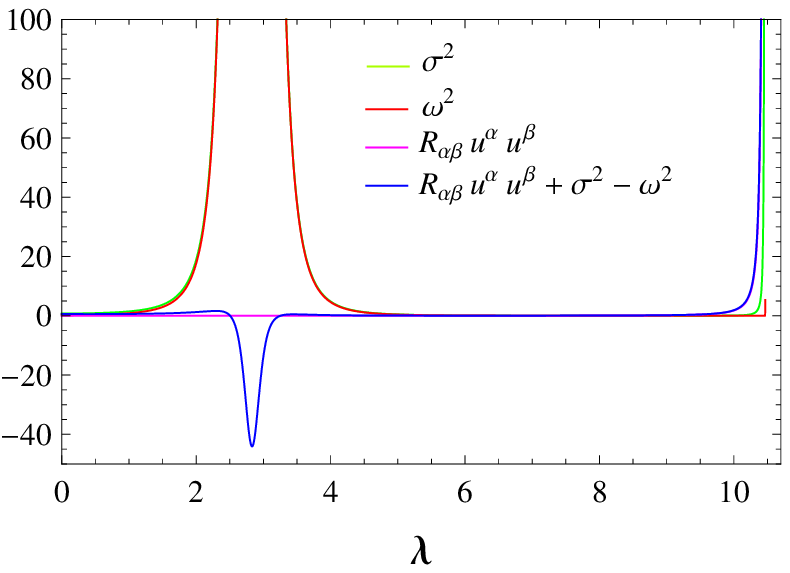}\label{fig:negative_alpha_rotation_2d}}
\caption{Plots of $R_{\alpha\beta}u^\alpha u^\beta$, $\sigma^2$, $\omega^2$ and $I$ ($=R_{\alpha\beta}u^\alpha u^\beta+\sigma^2-\omega^2$) for the initial values of the ESR variables corresponding to Fig. \ref{fig:negative_alpha_rotation_1}.}
\label{fig:negative_alpha_rotation_2}
\end{figure}

\noindent It is well known that rotation always  defies focusing. 
The $\sigma_0^2=constant$ sections of Fig. \ref{fig:negative_alpha_3D_3} show 
the dependence of $\lambda_f$ on initial rotation $\omega_0$. It is clear 
that $\lambda_f$ increases with $\omega_0^2$ up to a certain critical value 
$\omega_c^2$. At $\omega_0^2=\omega_c^2$, there is a sudden change in 
focusing time. To understand this transition, i.e., sudden change in 
focusing time, we plot $\theta(\lambda)$ for $\theta_0=0$ and 
$\sigma^2_0=0.75$ and note the change in focusing time for different 
initial rotation (Fig. \ref{fig:negative_alpha_rotation_1}). The corresponding 
plots for the ESR variables, $R_{\alpha\beta}u^\alpha u^\beta$ and $I$ are 
shown in Fig. \ref{fig:negative_alpha_rotation_2}. Clearly, one can note a 
sudden change in focusing time between $\omega^2_0=0.1731$ and 
$\omega^2_0=0.1735$. The transition takes place because of the dominance of 
$\omega^2(\lambda)$, midway during the evolution 
(Figs. \ref{fig:negative_alpha_rotation_2c} and  \ref{fig:negative_alpha_rotation_2d}).

\subsection{Evolution of kinematic variables for ${\bf \alpha=\frac{\sqrt{3}}{2}}$}
\begin{figure}[ht]
\centering
\subfigure[$\sigma^2_0=0.0$]{\includegraphics[scale=0.94]{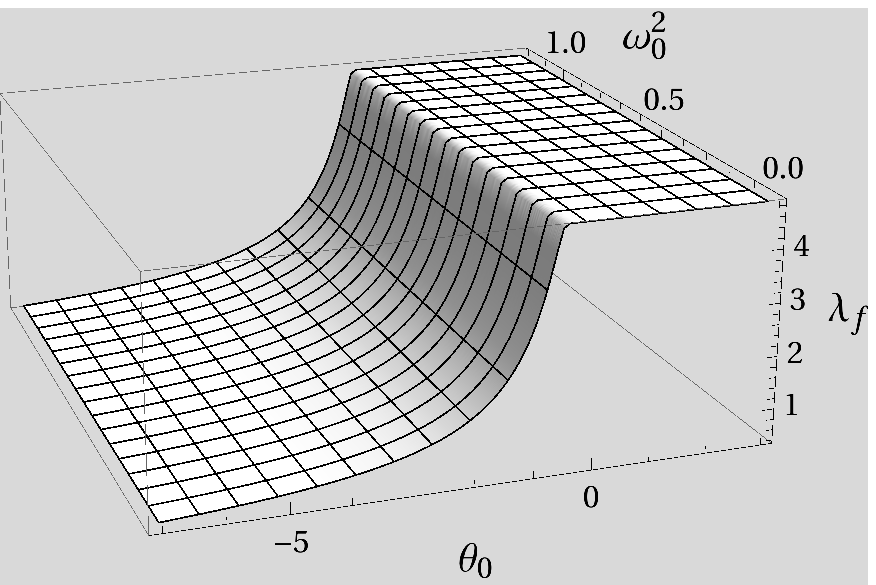}\label{fig:positive_alpha_3D_1}}\hspace{0.2cm}
\subfigure[$\sigma^2_0=0.35$]{\includegraphics[scale=0.85]{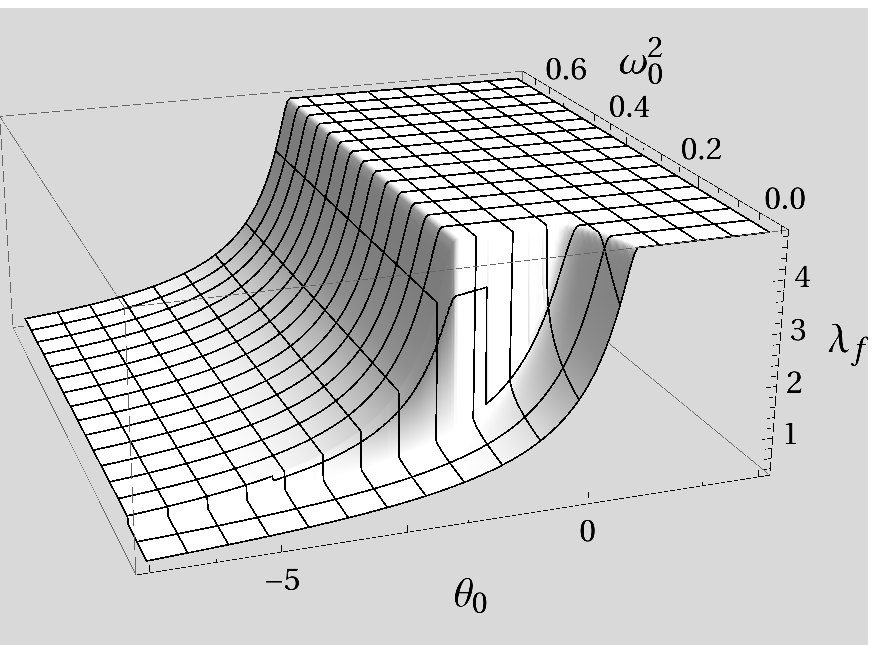}\label{fig:positive_alpha_3D_2}}
\subfigure[$\omega^2_0=0.0$]{\includegraphics[scale=0.89]{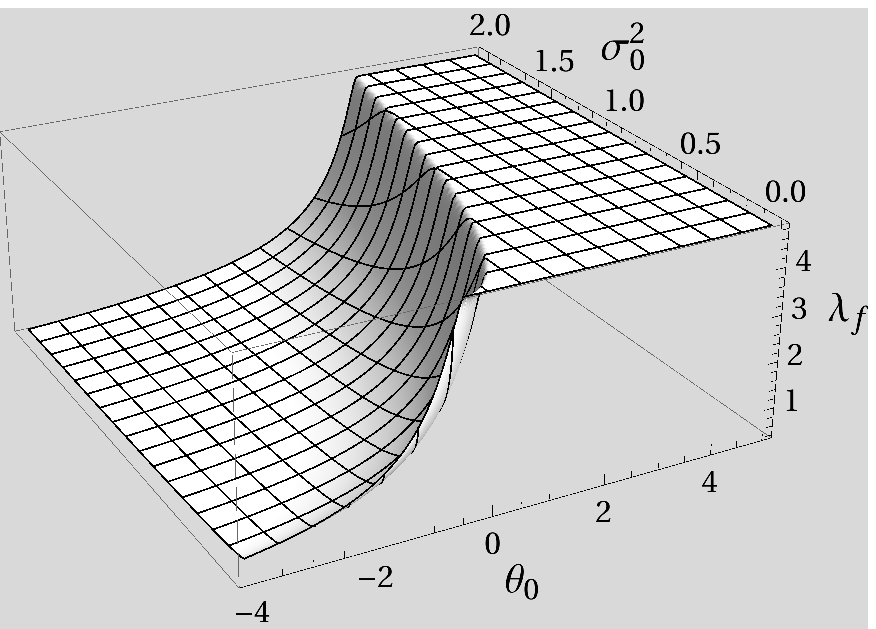}\label{fig:positive_alpha_3D_3}}\hspace{0.2cm}
\subfigure[$\theta_0=0.0$]{\includegraphics[scale=0.93]{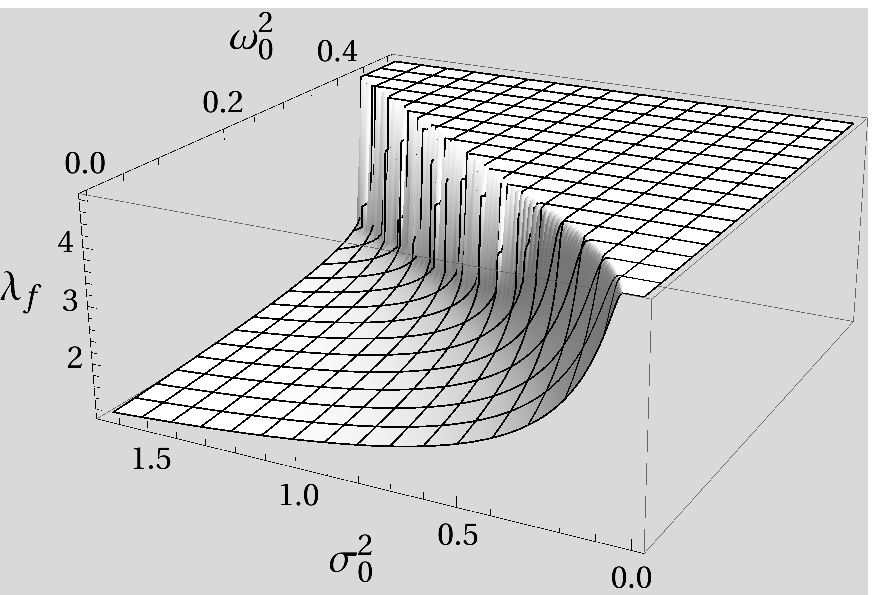}\label{fig:positive_alpha_3D_4}}
\caption{Plots showing the dependence of the focusing affine parameter $\lambda_f$ on the initial ESR for $\alpha=\frac{\sqrt{3}}{2}$. Here the initial conditions are $t(0)=-10.0$, $z(0)=0.3$, $\phi(0)=0$, $\dot{t}(0)=3.0$, and $L=4.0$.}
\label{fig:positive_alpha_3D}
\end{figure}

\noindent As in the previous case, here too focusing always takes place. 
In Fig. \ref{fig:positive_alpha_3D}, we show the dependence of the focusing 
affine parameter $\lambda_f$ on the initial values of the ESR variables. 
In this case, the singularity formation time $t=1.0$ corresponds to 
$\lambda_f=4.82$. 

\subsubsection{Effect of initial expansion and shear on focusing}
\noindent The $\omega_0^2=constant$ sections of Figs. \ref{fig:positive_alpha_3D_1} and \ref{fig:positive_alpha_3D_2} and $\sigma_0^2=constant$ sections of Fig. \ref{fig:positive_alpha_3D_3} indicate that, up to a certain initial 
expansion, focusing time increases with increasing $\theta_0$. Above this 
certain value, focusing time is independent of $\theta_0$ because focusing 
always takes place at the singularity. 
The dependence of focusing time on the initial shear is the same as that 
in the case with $\alpha=-\frac{\sqrt{3}}{2}$. 

\subsubsection{Effect of initial rotation on focusing}
\begin{figure}[ht]
\centering
\subfigure[$\theta_0=0.0,\sigma^2_0=0.0$]{\includegraphics[scale=0.75]{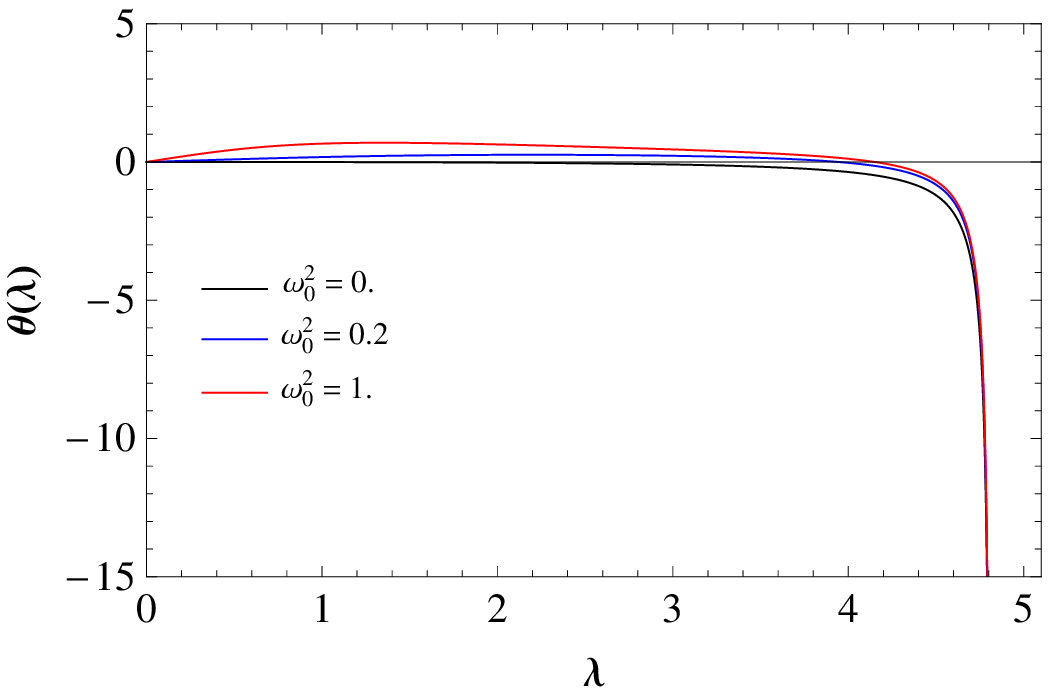}\label{fig:positive_alpha_rotation_1a}}
\subfigure[$\theta_0=0.0,\sigma^2_0=1.0$]{\includegraphics[scale=0.75]{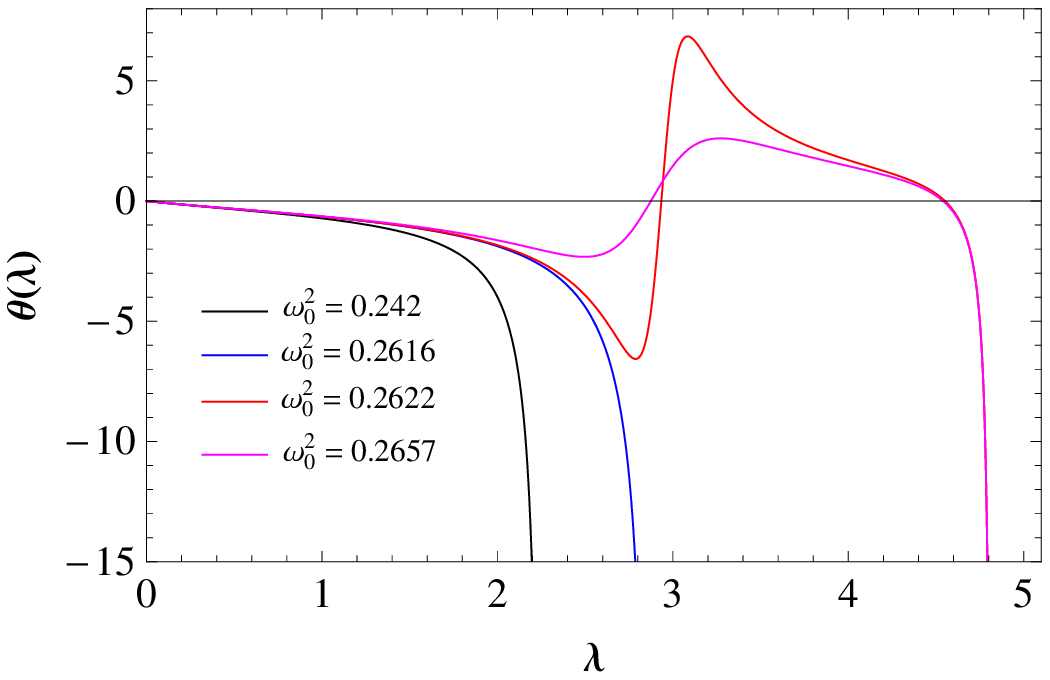}\label{fig:positive_alpha_rotation_1b}}
\caption{The role of the initial rotation $\omega^2_0$ on the evolution of the expansion scalar $\theta$ and 
focusing of the congruence with $\alpha=\frac{\sqrt{3}}{2}$.}
\label{fig:positive_alpha_rotation_1}
\end{figure}

\noindent The $\theta_0=constant$ sections of 
Fig. \ref{fig:positive_alpha_3D_1} indicate that, with zero initial shear, 
the focusing time is independent of rotation but with a sufficient nonzero 
initial shear, focusing time depends on the initial rotation 
[see $\sigma^2_0=constant$ sections of Fig. \ref{fig:positive_alpha_3D_4} 
and Fig. \ref{fig:positive_alpha_rotation_1b}]. This is due to the fact that, 
in the absence of any spacetime singularity formation, the congruence would 
have focused beyond the singularity formation time $t=1.0$. A sufficient 
initial shear focuses the congruence before it falls to the singularity. This 
focusing now can be delayed by choosing some initial rotation.
Thus, sufficient initial shear and rotation affect the focusing behavior of the congruence
[Figs. \ref{fig:positive_alpha_3D_4} and \ref{fig:positive_alpha_rotation_1b}].

\begin{figure}[ht]
\centering
\subfigure[$\theta_0=0.0,\omega^2_0=0.2$]{\includegraphics[scale=0.78]{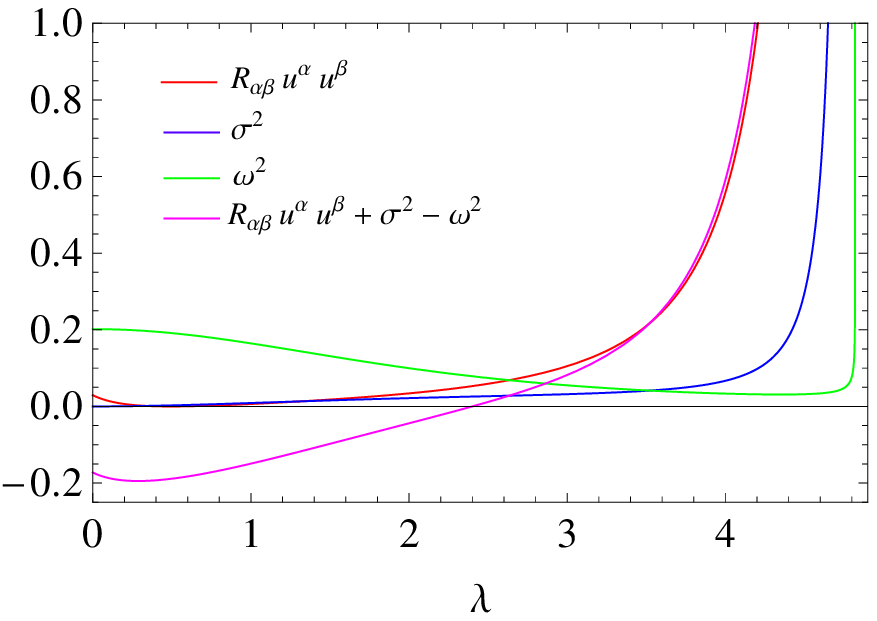}\label{fig:positive_alpha_rotation_2a}}\hspace{0.2cm}
\subfigure[$\theta_0=0.0,\omega^2_0=0.5$]{\includegraphics[scale=0.78]{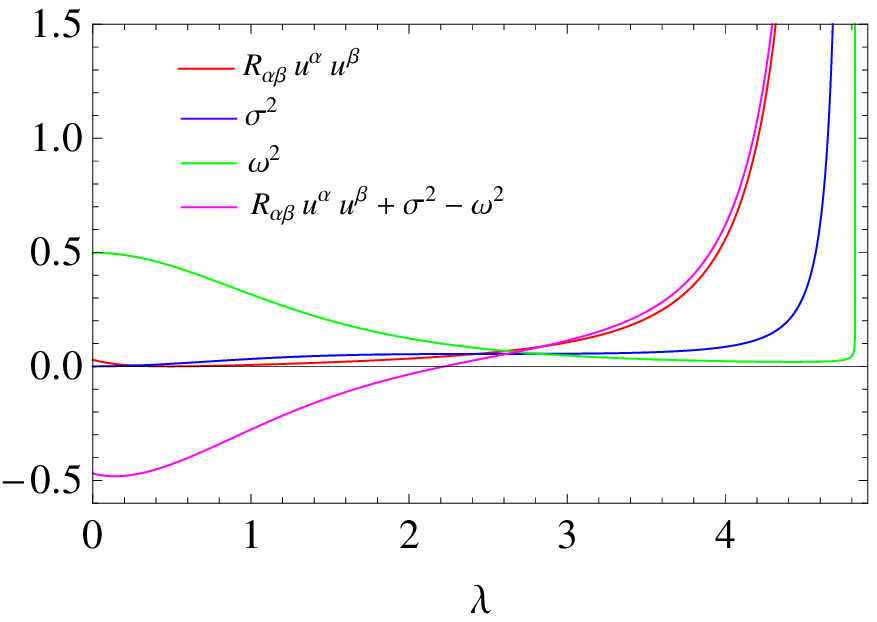}\label{fig:positive_alpha_rotation_2b}}
\subfigure[$\theta_0=-0.7,\omega^2_0=0.2$]{\includegraphics[scale=0.78]{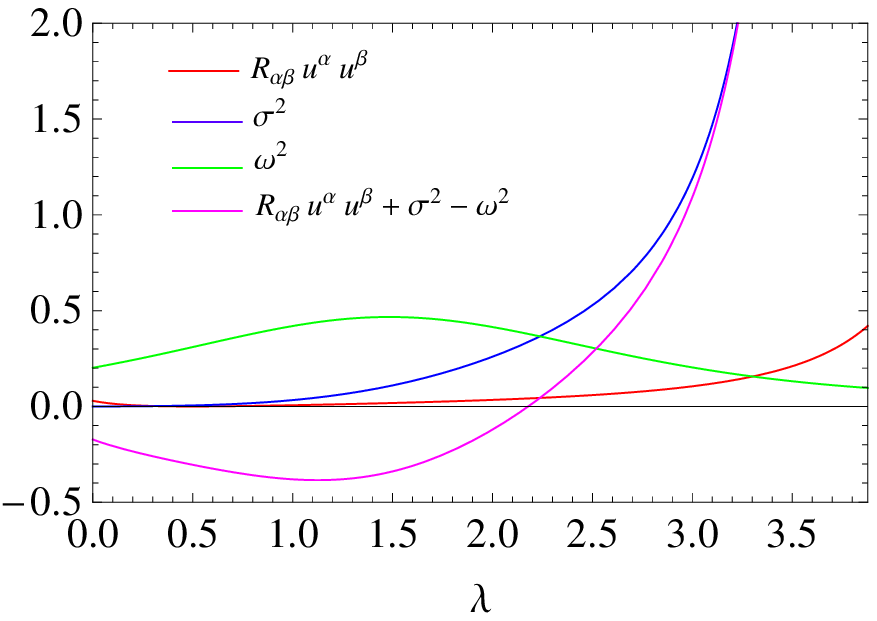}\label{fig:positive_alpha_rotation_2c}}\hspace{0.2cm}
\subfigure[$\theta_0=-0.7,\omega^2_0=0.5$]{\includegraphics[scale=0.75]{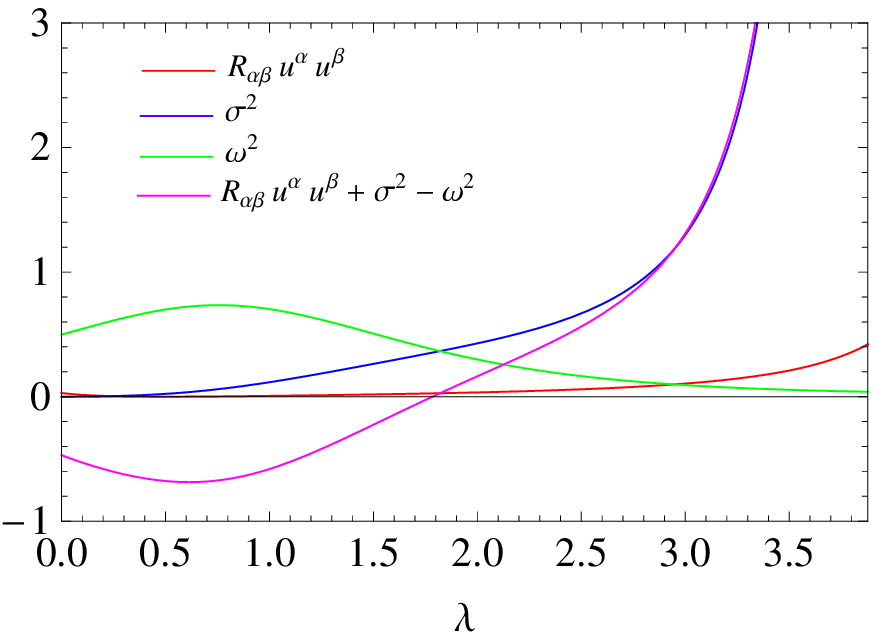}\label{fig:positive_alpha_rotation_2d}}
\caption{Plots of $R_{\alpha\beta}u^\alpha u^\beta$, $\sigma^2$, $\omega^2$ and $I$ ($=R_{\alpha\beta}u^\alpha u^\beta+\sigma^2-\omega^2$) 
for $\sigma_0^2=0.0$.}
\label{fig:positive_alpha_rotation_2}
\end{figure}

\noindent Figures \ref{fig:positive_alpha_rotation_2} demonstrates the reason 
that $\lambda_f$ is independent of initial rotation for $\sigma^2_0=0$. 
From the figures, it is clear that, unlike the case for 
$\alpha=-\frac{\sqrt{3}}{2}$, the rotation drops to small values and hence 
does not have a significant effect on $I$ as the evolution proceeds. 
Therefore, the evolution of $I$ is 
completely controlled either by the curvature term 
[Figs. \ref{fig:positive_alpha_rotation_2a} and \ref{fig:positive_alpha_rotation_2b}] or 
by the shear [Figs. \ref{fig:positive_alpha_rotation_2c} and \ref{fig:positive_alpha_rotation_2d}]. The transition in $\lambda_f$ in 
Fig. \ref{fig:positive_alpha_rotation_1b} can be explained in the same way 
as that for the case $\alpha=-\frac{\sqrt{3}}{2}$.

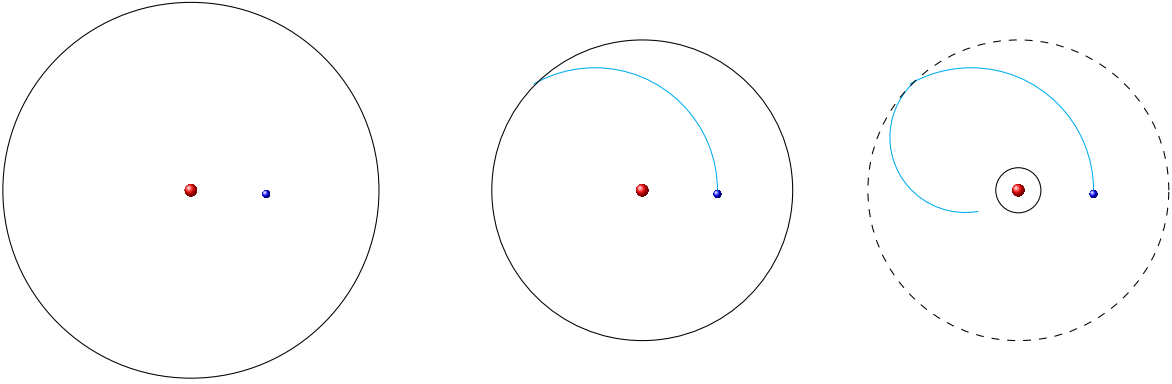
\begin{figure}[ht]
\centering
\begin{tikzpicture}
\centering
\shade[shading=ball,ball color=red] (0,0) circle (0.08);
\shade[shading=ball,ball color=blue] (1,-0.05) circle (0.05);
\draw (0,0) circle (2.5);
\shade[shading=ball,ball color=red] (6.0,0) circle (0.08);
\shade[shading=ball,ball color=blue] (7.0,-0.05) circle (0.05);
\draw[cyan] (7.0,0) arc [radius=1.63, start angle=0, end angle= 120];\draw (6.0,0) circle (2.0);
\draw[cyan] (12,0) arc [radius=1.63, start angle=0, end angle= 120];\draw[cyan] (9.65,1.47) arc [radius=1.0, start angle=130, end angle=280];\draw[dashed] (11,0) circle (2.0);\draw (11,0) circle (0.3);
\shade[shading=ball,ball color=red] (11.0,0) circle (0.08);
\shade[shading=ball,ball color=blue] (12.0,-0.05) circle (0.05);
\end{tikzpicture}
\caption{Schematic diagrams explaining the focusing of congruence. The solid circle indicates the (shrinking) apparent horizon and the blue curve traces the central geodesic of the congruence.}
\label{fig:schematic}
\end{figure}

\paragraph*{}
From the above analysis, we observe that focusing always takes place. To 
illustrate this further, we draw three schematic diagrams  in
Fig. \ref{fig:schematic}. The solid circle, red dot, and blue dot represent the 
position of the apparent horizon, the singularity at $z=0$, and the initial 
position of the congruence, respectively. The region outside (inside) the 
apparent horizon is trapped (untrapped). The apparent horizon shrinks in size with 
time. As the evolution proceeds, depending on the initial conditions on the ESR variables, the congruence may get focused before hitting the apparent horizon or may hit the apparent horizon at some $\lambda$ [Fig. \ref{fig:schematic}] and get trapped. This trapped congruence gets focused either before falling into the singularity or at the singularity (third diagram). Therefore, we must always have focusing, though this 
may be benign (not happening at a curvature singularity) for certain initial conditions,
as discussed above.

\section{Analysis of jump in the expansion scalar}
\label{analysis}

\noindent In the last section, we noticed that midway during the evolution
of the congruence, the dominance of rotation over shear (which makes $I<0$) 
leads to a sharp transition (from negative to positive value) in the evolution of expansion of the congruence. As the 
evolution proceeds further, because of the curvature term, $I$ diverges 
to positive infinity, thereby causing eventual focusing of the congruence. 
Let us first ask and analyze this question: what would happen if the curvature term does 
not diverge as the evolution proceeds? For example, for static spacetimes, 
as the family of outgoing timelike geodesics evolves, the curvature term 
$R_{\alpha\beta}u^\alpha u^\beta$ becomes less and less significant. This is 
also true for this nonstatic case, except at the end of the evolution process. 
Therefore the value and sign of $I$ is largely determined by the values of the
shear and the rotation terms.  
Ignoring the curvature term, we may have the following subcases:

\noindent (1) During the evolution, if there is a phase
where $I<0$ (rotation dominates over shear) and then the shear dominates over 
the rotation, ultimately making $I> 0$, we may observe a finite jump
after which the congruence will focus.

\noindent (2) If $I < 0$ always and goes to zero as $\lambda \to \infty$, 
$\theta$ goes to zero after a
sharp transition from negative to positive value. This effect 
has been observed earlier in 
the kinematic study of a family of projectile trajectories \cite{rajab} 
and also in the kinematic study of deformable media without stiffness 
\cite{ADG3}.

\noindent (3) If $I$ oscillates between positive and negative values, 
we have periodic oscillations in the expansion scalar; focusing does not take 
place.  
This effect has been observed in the kinematic study of a family of 
trajectories in the two dimensional 
isotropic harmonic oscillator, a charged particle in an 
electromagnetic field \cite{rajab} and also in the kinematic study of 
deformable media with stiffness \cite{ADG3}. 
In \cite{rajab}, $I=\sigma^2-\omega^2$ and the analog of the 
curvature term is a constant, given by $\alpha$; the total $I+$curvature
oscillates between positive and negative values.

\noindent (4) If $I < 0$ always and diverges to negative infinity at some 
$\lambda$, we may have complete defocusing of the congruence.

\noindent One of the above cases could have occurred for our spherically symmetric, nonstatic
spacetime, had the singularity not formed but as the 
evolution proceeds toward the singularity formation time, the curvature term, 
and hence $I$, diverges to positive infinity. Therefore, we have focusing 
following a sharp jump in the expansion scalar. If the rotation had not
dominated over shear, midway during the evolution, we would have observed 
direct focusing without any intermediate jump in the expansion scalar.

\begin{figure}[ht]
\centering
\includegraphics[scale=0.50]{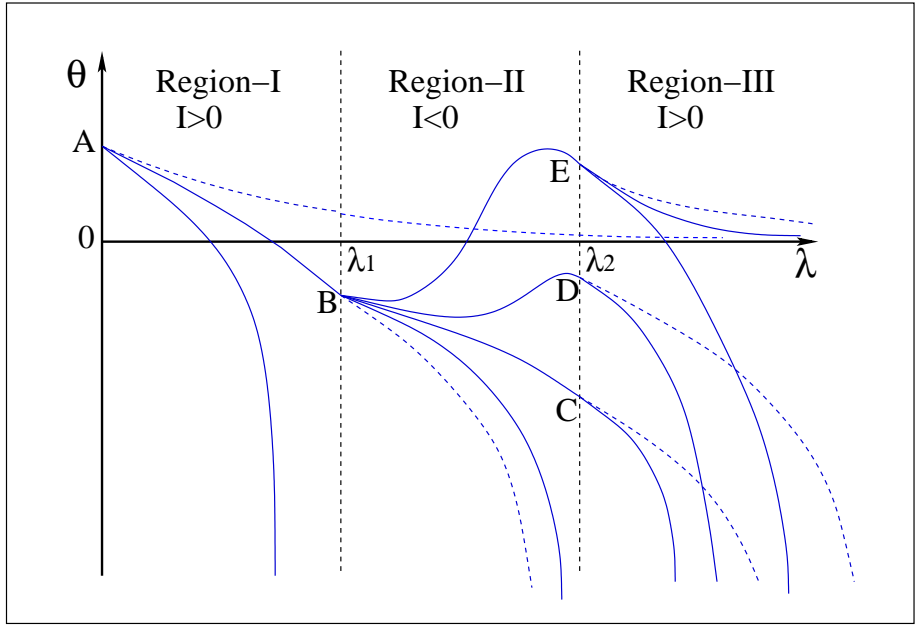}
\caption{Schematic diagram showing the possible evolution of the expansion (solid blue curves) when $I(\lambda)$ becomes negative over a finite domain of $\lambda$. $I(\lambda)$ is positive in regions I and III and is negative in region II. $I(\lambda)=0$ at $\lambda=\lambda_1$ and $\lambda=\lambda_2$. The dotted blue curves show the evolution of expansion for identically vanishing $I$. See the discussion in Sec. \ref{analysis}.}
\label{fig:theta_lambda}
\end{figure}

\noindent We now explain this nonmonotonic glitch/jump in the expansion 
before complete focusing in a general way. To do this, let us go back to
the Raychaudhuri equation for the expansion given as
\begin{equation}
\frac{d\theta}{d\lambda} +\frac{1}{3}\theta^2 = -I(\lambda)
\end{equation}
Let us assume that $I(\lambda)$ is negative between $\lambda=\lambda_1$
and $\lambda=\lambda_2$ and positive elsewhere (as shown in Fig. \ref{fig:theta_lambda}). What can we 
say about the possible behavior of $\theta$ in such a case? When $I(\lambda)\geq 0$ (region I in Fig. \ref{fig:theta_lambda}), we have
\begin{equation}
\frac{d\theta}{d\lambda} +\frac{1}{3}\theta^2 \leq 0.
\end{equation}
Integrating the above inequality gives 
\begin{equation}
\theta (\lambda) \leq \frac{1}{\frac{\lambda}{3} +\frac{1}{\theta_0}}
\end{equation}
where $\theta_0$ is the value of $\theta$ at $\lambda=0$.
Hence, any expansion curve $\theta(\lambda)$  (with $I>0$) in this region I 
must obey the above inequality. The slope of $\theta(\lambda)$ 
at any $\lambda$ must
also be less than the slope of the equality curve for the expansion. The equality curve, i.e., the expansion curve for $I=0$, is shown by dotted blue curve. 
It is possible that
the expansion curve for $I>0$ diverges to negative infinity within this region I,
thereby resulting in focusing. If it does not focus, it will hit the
$\lambda=\lambda_1$ line at some value of $\theta$ (at point B, say). Taking this value as the
initial $\theta$, one now proceeds into region II where $I<0$. In this
region the inequality reads to
\begin{equation}
\theta (\lambda) \geq \frac{1}{\frac{\lambda-\lambda_1}{3} +\frac{1}{\theta_{\lambda_1}}}
\label{eq:inequality2}
\end{equation}
where $\theta_{\lambda_1}$ is the value of the expansion at
$\lambda_1$ for some typical expansion curve in region I ending
at $\lambda_1$.
It is now possible for the expansion curve to turn around
and enter the region II allowed
by the inequality \ref{eq:inequality2} in region II. In this region, if $I(\lambda)$ is not negative enough so that we always have $\frac{d\theta}{d\lambda}<0$, then the curve may either get focused in this region or hit region III at $\lambda=\lambda_2$ (at point C, say). However, if $I(\lambda)$ is negative enough, then $\frac{d\theta}{d\lambda}$ can change sign and become positive. Eventually,
the expansion curve reaches $\lambda_2$ (at points D or E, say) with a certain value of $\theta$,
 which is
taken as the initial value for region III. Drawing the equality curve (blue dotted curve) in
region III with
this initial value, one obtains the corresponding allowed part in region III. 
The expansion curve can therefore either focus or do what it did in region I. It may be noted that the equality curve in region II for the negative initial expansion may hit region III at some negative $\theta$; in that case, the expansion curve for negative $I(\lambda)$ will not focus in region II and will always hit region III.
The existence of region II with $I<0$, between the two regions I and III where
$I>0$, creates the jump in the expansion before it diverge to negative
infinity. At $\lambda=0$, if we start with a negative expansion, we find similar possibilities for 
the expansion curve.

\noindent Thus, the cause behind the jump in the expansion is explained
using the fact that the dominance of rotation over shear plus curvature
can lead to a domain in $\lambda $ where $I(\lambda)$ is negative.
The sandwiching of this region between those with positive $I(\lambda)$
results in the sudden jump. If more such regions exist, it is evident that
the expansion may exhibit a repetition of such behavior before
focusing. 

\section{Effect of Nonstaticity and inhomogeneity}
\label{metric_parameter}
\begin{figure}[ht]
\centering
\subfigure[$a=0.0,c=0.0$]{\includegraphics[scale=0.80]{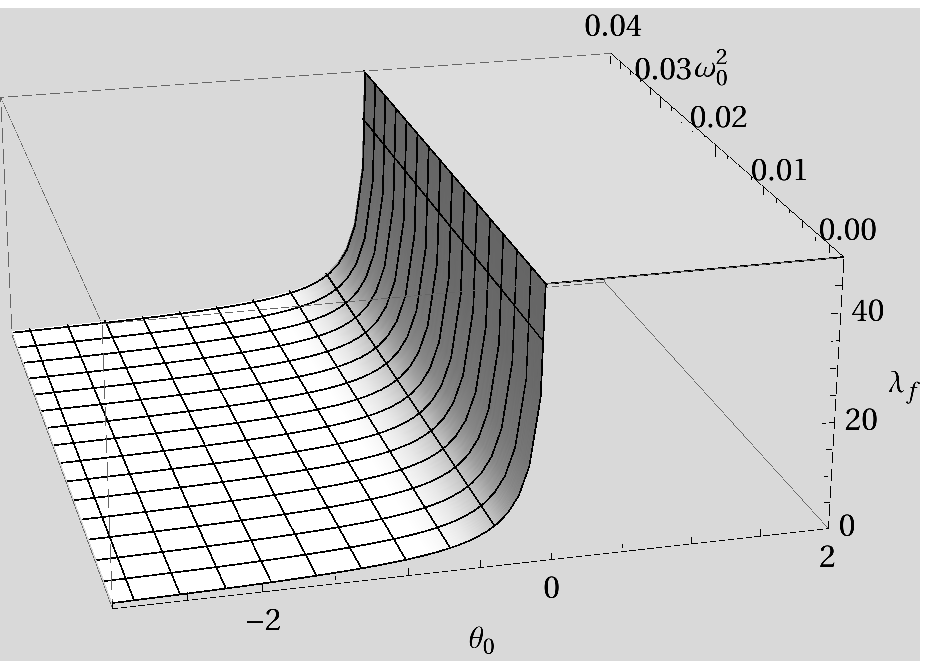}\label{fig:new3D_1}}\hspace{0.2cm}
\subfigure[$a=0.0,c=1.0$]{\includegraphics[scale=0.77]{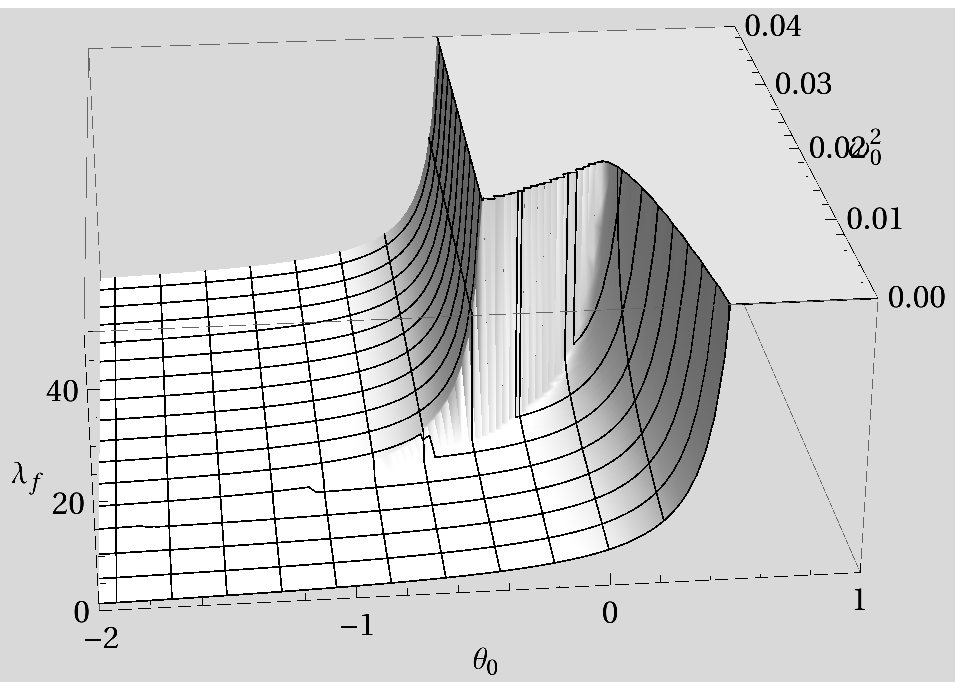}\label{fig:new3D_2}}
\subfigure[$a=-0.01,c=0.0$]{\includegraphics[scale=0.85]{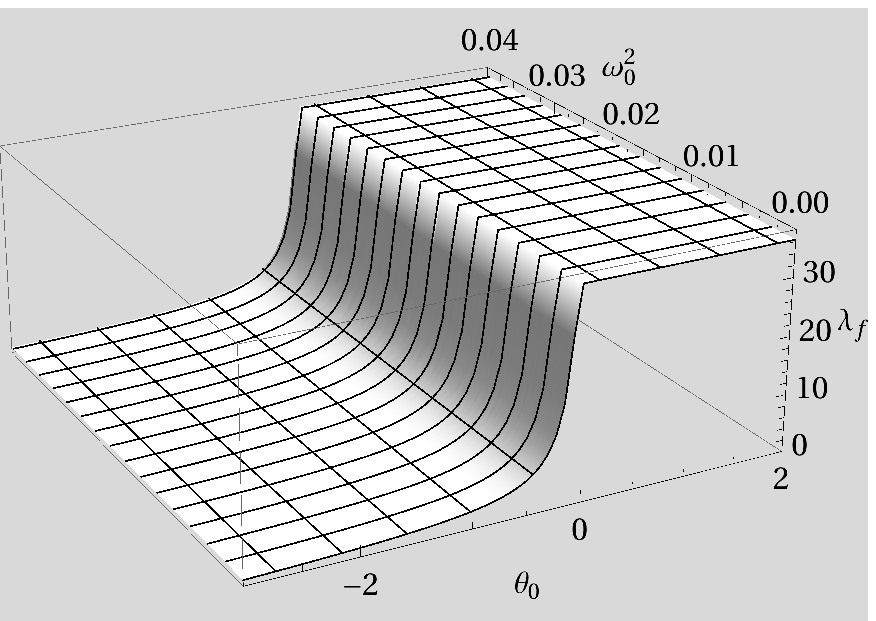}\label{fig:new3D_3}}\hspace{0.2cm}
\subfigure[$a=-0.01,c=1.0$]{\includegraphics[scale=0.71]{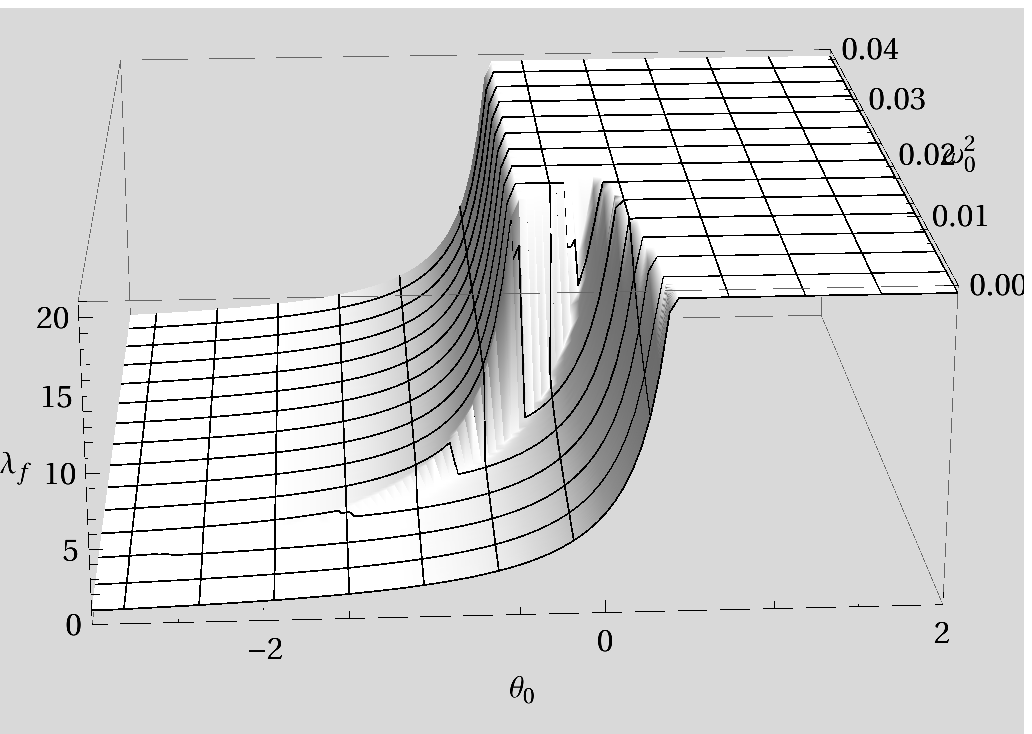}\label{fig:new3D_4}}\hspace{0.2cm}
\subfigure[$a=-0.01,c=8.0$]{\includegraphics[scale=0.87]{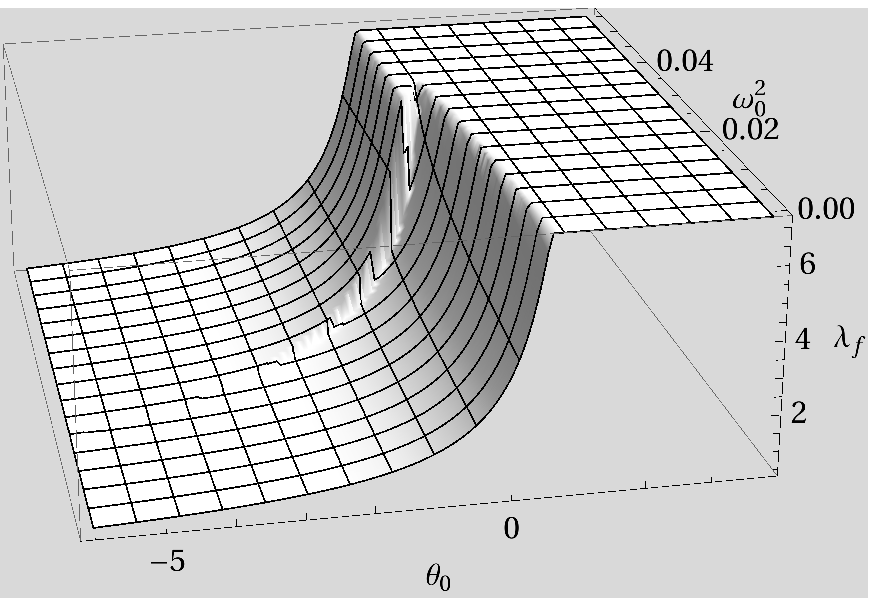}\label{fig:new3D_6}}\hspace{0.2cm}
\subfigure[$a=-0.01,c=20.0$]{\includegraphics[scale=0.83]{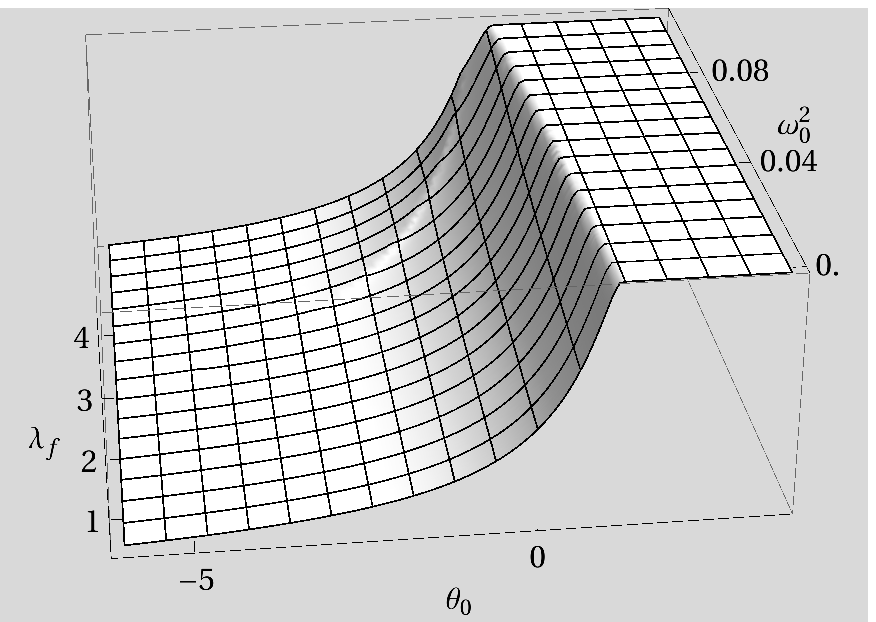}\label{fig:new3D_10}}
\caption{Plots showing the dependence of the focusing affine parameter $\lambda_f$ on the initial expansion and rotation for $\alpha=-\frac{\sqrt{3}}{2}$, $t(0)=0.0$, $z(0)=2.0$, $\phi(0)=0$, $\dot{t}(0)=2.0$, and $L=2.2$. Here $\sigma_0^2=0.0$.}
\label{fig:new3D}
\end{figure}

\begin{figure}[ht]
\centering
\subfigure[$a=0.0,c=0.0$]{\includegraphics[scale=0.67]{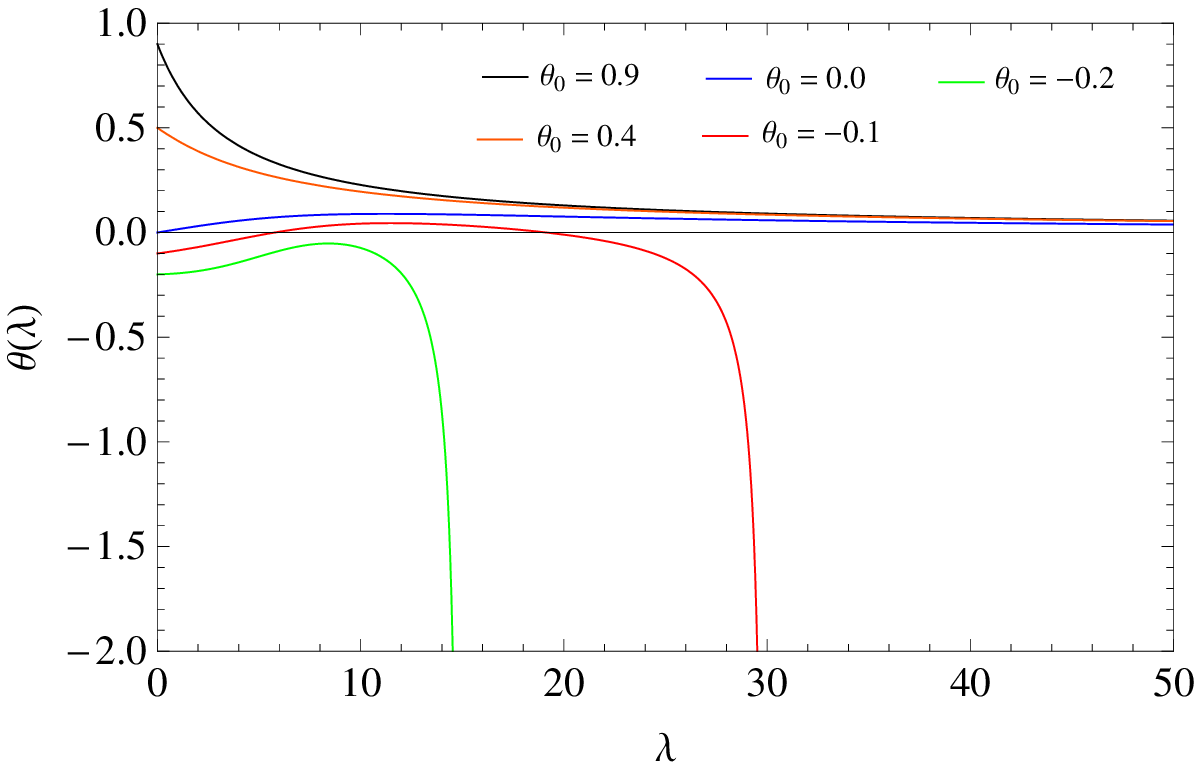}\label{fig:new2D_1}}\hspace{0.2cm}
\subfigure[$a=0.0,c=1.0$]{\includegraphics[scale=0.67]{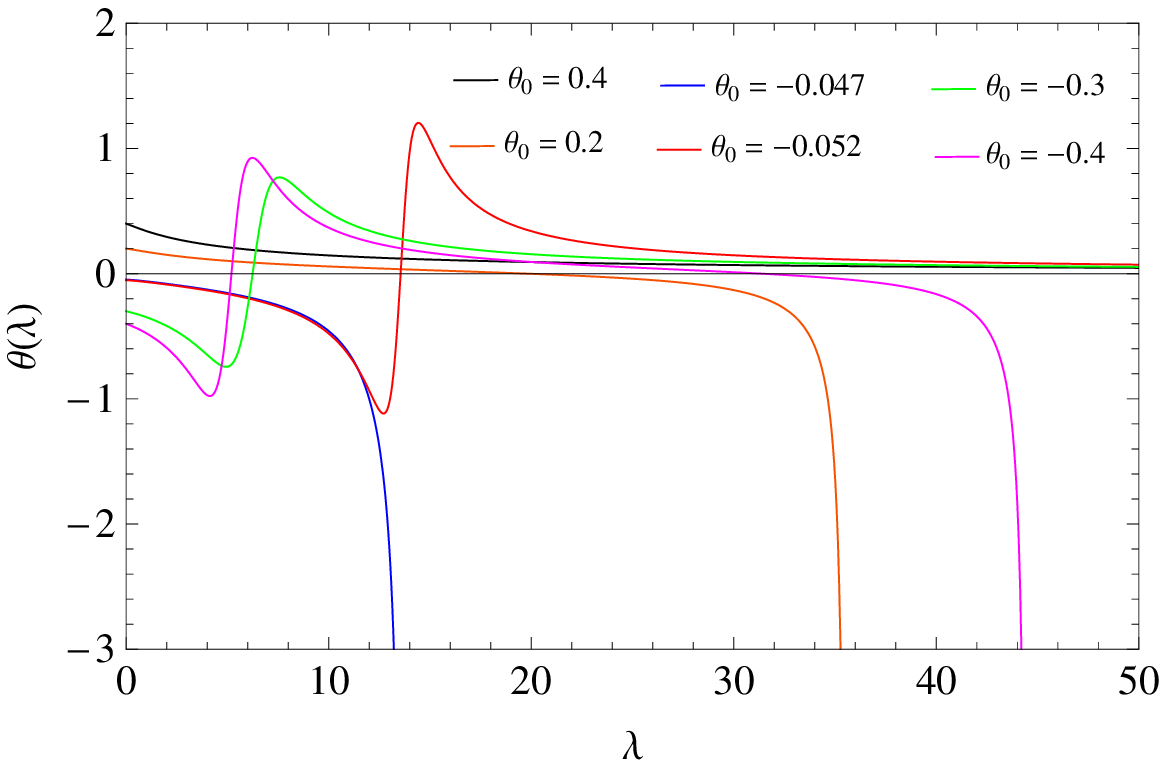}\label{fig:new2D_2}}
\subfigure[$a=-0.01,c=0.0$]{\includegraphics[scale=0.67]{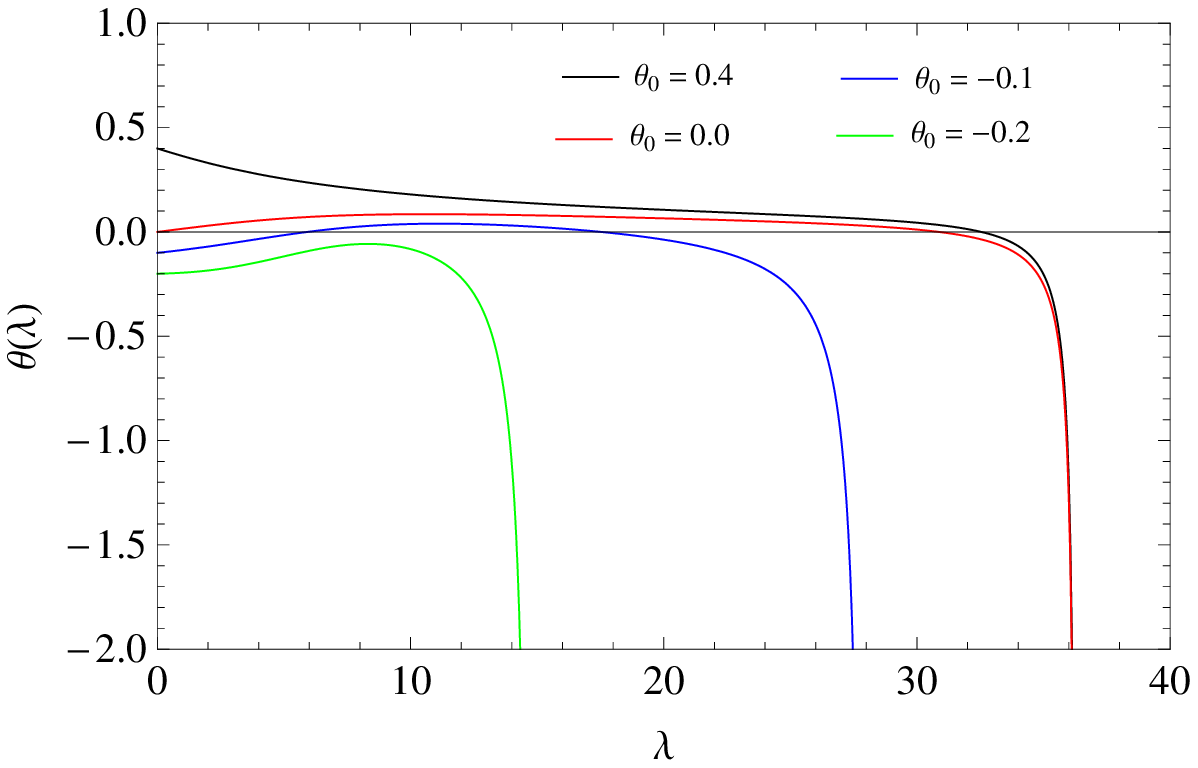}\label{fig:new2D_3}}\hspace{0.2cm}
\subfigure[$a=-0.01,c=1.0$]{\includegraphics[scale=0.67]{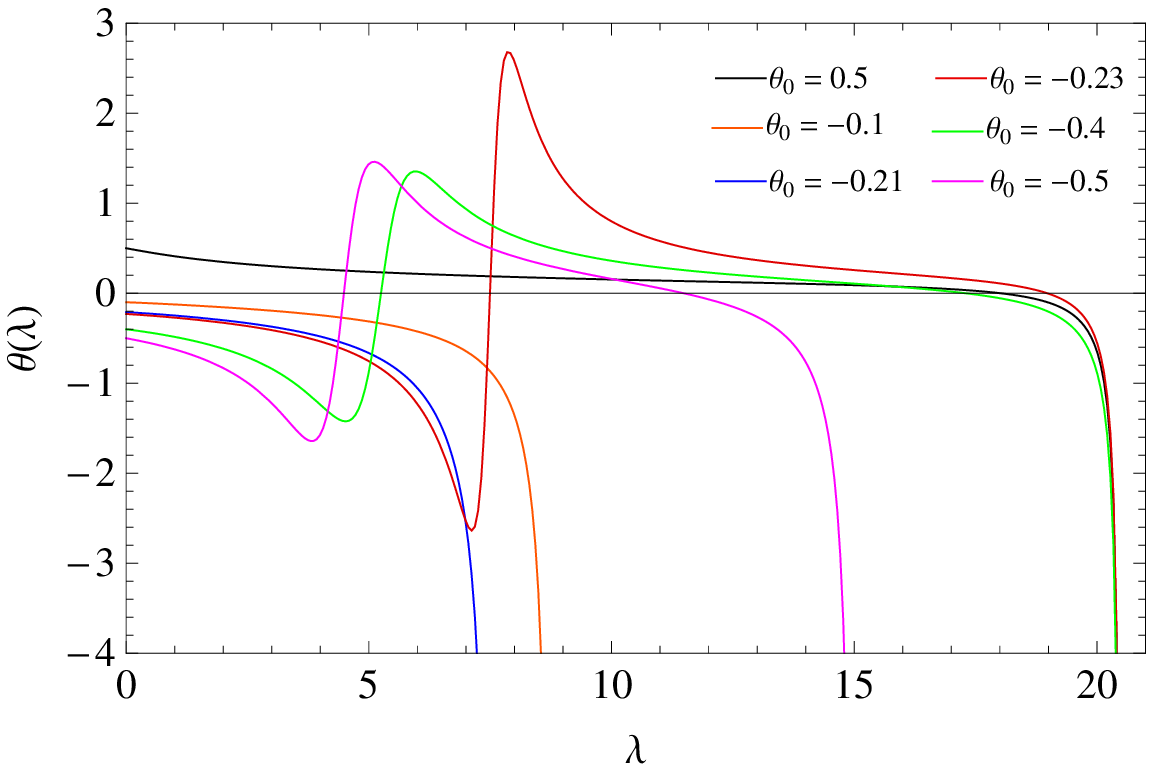}\label{fig:new2D_4}}
\caption{Plots showing the evolution of expansion for the initial conditions in Fig. \ref{fig:new3D}. Here $\sigma_0^2=0.0,\omega^2_0=0.016$.}
\label{fig:new2D}
\end{figure}

\noindent 
Given the analysis of congruences in the previous sections, we may
ask the following question.
Which among the two parameters in the line element (i.e., $a$ and $c$)
is responsible for the jump in the expansion before complete focusing?
To see this, in this section, we study the effect of $a$ and $c$ on the 
kinematic evolution of geodesic congruences. The central 
inhomogeneity parameter $c$ plays an important role in the evolution of 
the timelike congruences. For the initial 
conditions in Fig. \ref{fig:negative_alpha_3D}, the timelike geodesic equations cannot be solved in 
the limit $c \to 0$ or $a \to 0$. Therefore, we take different initial 
conditions on 
$\lbrace x^\alpha(\lambda),u^\alpha(\lambda)\rbrace$, which can be used to solve the timelike geodesic 
equations for all values of $c$ and $a$. The dependence of the time to 
singularity, quantified by 
$\lambda_f$, on the initial expansion and rotation for different values of 
$a$ and $c$ is shown in 
Fig.~\ref{fig:new3D}.  The sharp rise of $\lambda_f$ in Figs. \ref{fig:new3D_1} and \ref{fig:new3D_2} 
indicates that $\lambda_f \to \infty$; i.e., focusing does not occur. 
Therefore, in the static case ($a=0$),
irrespective of the presence/absence of the central inhomogeneity $c$, there 
exists a region in the space of initial conditions for which focusing does not takes place; the expansion scalar goes 
to zero [Fig. \ref{fig:new2D_1} and Fig. \ref{fig:new2D_2}] as $\lambda \to \infty$. On the other 
hand, in the collapsing case ($a\neq 0$), we always have focusing in finite time. 
Interestingly, the presence of the central inhomogeneity seems to induce 
the  above-mentioned jump in the evolution of the expansion scalar (Fig. \ref{fig:new2D}). 
From Figs. \ref{fig:new3D_3}-\ref{fig:new3D_10}, 
it is clear that the jump behavior is absent for $c=0$ and also for large value of $c$. From 
the geodesic equations, we notice that, for a given initial condition, $\dot{\phi}\sim (z+2c)^{-(1+\alpha)}$. 
Therefore, for large $c$, $\dot{\phi}\approx 0$, which means that the congruence is almost radial. 
Thus, the burst in congruence rotation, as observed for moderate values of $c$, is damped out. 
This in turn smooths out the jump in the evolution of the expansion scalar
which may be the reason behind the gradual disappearance of the peculiar
behavior in the variation of
the time to singularity with initial conditions, as observed in Figs.~\ref{fig:new3D_4}-\ref{fig:new3D_10}. In essence, the inhomogeneity in the line element
is responsible for the jump in the expansion before focusing. The
nonstaticity does not lead
to any qualitative changes in the jump phenomenon.


\section{Summary and Conclusion}
\label{summary_conclusion}

\noindent In this work, we have demonstrated how the distinct roles of
shear, rotation, and spacetime curvature affect the detailed behavior
of trajectories in a gravitational scalar field collapse scenario. 
The dependence of the time to singularity $\lambda_f$ on the initial conditions has been spelled out
in detail.  The difference in the nature of the spacetime geometry for
the two different values of $\alpha$ is shown to be reflected in
the quantitative behavior of the congruence of geodesics. 
We summarize the key findings as follows.
\begin{itemize}
\item For the cases in which a spacelike curvature singularity is formed, a (timelike) congruence 
is eventually trapped and focuses either by hitting the curvature singularity or due to
intersection of geodesics in finite time. In the latter case, a peculiar influence of the initial expansion and rotation
on the time to singularity is observed. 

\item For initial conditions on the kinematic variables in a certain range, the occurrence of congruence 
singularity due to intersection of geodesics in finite time is observed to be delayed. It is
found that the expansion scalar exhibits a jump from negative (contracting) to positive (expanding) before it focuses eventually. Using numerical 
results, we have seen that the jump is triggered by a
burst in the rotation of the congruence. We have analyzed in detail
and in quite some amount of generality when and
why such a nonmonotonic  behavior in the expansion scalar can occur.   

\item For a large negative value of the initial expansion (initially rapidly collapsing congruence), the 
congruence singularity is driven by buildup of shear. The congruence singularity occurs much earlier than the time
taken by the central geodesic to hit the curvature singularity.

\item The central inhomogeneity parameter $c$ is observed to influence the geodesic behavior and, for moderate values, 
introduces a peculiar jump (burst) behavior in the evolution of the expansion (rotation) of the congruence. However, for high values of
$c$, the geodesics tend to be radial, which damps out the observed burst in 
congruence rotation. This in turn
smooths out the peculiar jump in the evolution of the expansion scalar.
In a way, therefore, the inhomogeneity is a cause behind the jump. 
For homogeneous spacetimes (i.e., when $c=0$), there is no such jump
in the expansion.


\end{itemize}

\noindent We have chosen a very specific exact solution for our
studies. The solution, by no means, represents a realistic collapse
scenario. However, the advantage of using this solution is
related to its exact nature, which is indeed rare, especially 
for space- and time-dependent (i.e., nonstatic) line elements. We
hope to study more realistic collapse scenarios in our future
investigations. 

\section*{Acknowledgment}
\noindent R. S. acknowledges the Council of Scientific and Industrial Research, India for providing support through a fellowship.

\end{document}